\documentclass[twocolumn,showpacs,amsmath,amssymb,pra]{revtex4}

\usepackage{graphicx}
\usepackage{dcolumn}
\usepackage{bm}
\usepackage{subfigure}

\begin{document}


\title {
Landau diamagnetic response in metals as a Fermi surface effect
}

\author {A. V. Nikolaev}

\affiliation{
Skobeltsyn Institute of Nuclear Physics, Moscow
State University, Vorob'evy Gory 1/2, 119234, Moscow, Russia
}

\affiliation{Department of Problems of Physics and Energetics, Moscow Institute of Physics and Technology, 141700 Dolgoprudny, Russia}

\date{\today}

\begin{abstract}
It is demonstrated that the Landau diamagnetism of the free electron gas and a monovalent metal can be considered as a Fermi surface effect.
Only a relatively small number of electron states close to the Fermi surface are diamagnetically active whereas
the majority of the electron states inside the Fermi surface are diamagnetically inert.
This partitioning of the occupied electron states is driven by the structure of Landau levels, around which one
can introduce magnetic tubes in the reciprocal space. Completely filled magnetic tubes do not change their energy in
an applied magnetic field, and only partially occupied magnetic tubes in the neighborhood of the Fermi surface
exhibit a diamagnetic response.
Using this partitioning of the occupied electron states we derive a general expression for the steady diamagnetic
susceptibility, for calculation of which one needs to know the shape of the Fermi surface and the energy gradient on it.
The method is applied to alkali metals, whose Fermi surfaces and energy gradients have been obtained from
{\it ab initio} band structure calculations.
It has been found that the Landau diamagnetic susceptibility is anisotropic depending on
the direction of the applied magnetic field with respect to the Fermi surface. This effect is more pronounced
for Li and Cs, whose Fermi surfaces show a noticeable deformation from the spherical shape.
The method opens a new route for {\it ab initio} calculations of the Landau diamagnetism of metals
or intermetallic compounds.
In the case of free electron gas it is shown that this approach also fully describes the oscillatory
de Haas - van Alphen part of the diamagnetic susceptibility.
Small oscillations of the Fermi energy found in the model are caused by redistribution (inflow or outflow) of electrons
from the equatorial region of the Fermi surface.
\end{abstract}



\maketitle

\section{Introduction}
\label{sec:int}

When an external magnetic field $H$ is applied to a metallic solid it induces a closed (orbital) motion of the itinerant electrons
resulting in a net nonvanishing magnetization antiparallel to $H$, which we call the Landau diamagnetism \cite{Lan0}.
In general the Landau diamagnetic susceptibility $\chi_L$ of solids is small and independent of the
temperature. In addition to the steady diamagnetism, at very low temperatures and strong magnetic fields
there are oscillatory dependencies ($\chi_{dHvA}$) due to the well-known de Haas - van Alphen effect \cite{Ons,Lif,Sho,AM}.

Despite many efforts in the past, an effective method for {\it ab initio} calculations of the Landau diamagnetism even
within the single-electron paradigm has not been established \cite{Cal,Mis}.
(As summarized in Ref.\ \cite{Cal}: ``The problem of the low field susceptibility of electrons is old,
interesting, and, unfortunately, quite complicated".)
This is because the effect is very difficult from the technical point of view.
Even the derivation of this effect for the free electron gas formulated by Landau \cite{Lan0} is rather complex.

In his pioneer work Peierls \cite{Pei} was the first who extended the Landau treatment to the case of electron band energy law $E(\vec{k})$
within the tight binding approximation.
(Alternative derivations were proposed by Wilson \cite{Wil1}, Hebborn and Sondheimer \cite{Heb1}, and recently by Briet {\it et al.} \cite{Bri}.)
Peierls obtained an expression for the magnetic susceptibility consisting of three terms,
\begin{eqnarray}
     \chi = \chi_1 + \chi_2 + \chi_3 ,
\label{i1}
\end{eqnarray}
where $\chi_1$ is the diamagnetic susceptibility analogous to that of isolated metal atoms,
$\chi_2$ is a term which has no simple physical interpretation and the term $\chi_3$ at zero temperature is given by
\begin{eqnarray}
 \chi_3 = -\frac{e^2}{48 \pi^3 \hbar^2 c^2} \int
 \left\{ \frac{\partial^2 E}{\partial k_x^2} \frac{\partial^2 E}{\partial k_y^2}
   - \left( \frac{\partial^2 E}{\partial k_x \partial k_y}  \right)^2 \right\} \frac{dS}{\nabla_k E} .  \nonumber \\
 \label{i2}
\end{eqnarray}
Here the integration is taken over the Fermi surface.
The term $\chi_3$ was considered as leading in the diamagnetism of conduction electrons \cite{Pei,Wil1,Wil2}.
For a simple band it reduces to the Landau-Peierls expression where in the equation for free electron susceptibility
the electron mass is replaced by an effective mass $m^*$.
In principle, Eq.\ (\ref{i2}) suggests that the diamagnetic effect is due to the Fermi surface electron states.
However, this statement can not be proved or considered rigorous because of the other contributions (like $\chi_1$, $\chi_2$ in Eq.\ (\ref{i1})) \cite{Pei,Wil1,Wil2,Heb1}.
In Ref.\ \cite{Wil2} on the basis of the study of the density matrix in the magnetic field Wilson, presenting a refined derivation of
Eq.\ (\ref{i2}), has found that some terms can not be explicitly evaluated and
some are not expressible in terms of derivatives of the band energy.
More reservations were added by Adams \cite{Ada} who claimed that the Landau-Peirels susceptibility was not always
the dominant contribution. Kjeldaas and Kohn \cite{KK} further suggested that the Landau-Peierls approximation is valid only
in the limit of small electron density. This statement has been proven recently by Briet {\it et al.} \cite{Bri}
at the mathematical level of accuracy. They have also worked out all satellite terms which in general accompany
Eq.\ (\ref{i2}) but disappear for small electron density.

The full quantum treatment of band diamagnetism in a real solid is a formidable task for a number of reasons.
First, the electron motion in the direction of the magnetic field $H$ differs from
the motion in the perpendicular plane. In a solid on the other hand, we deal with the translational symmetry
which equally holds in three dimensions. Therefore, on applying the magnetic field to a solid, its translational symmetry
becomes broken, and accurate description of electron bands and even their classification within the former first Brillouin zone is impossible \cite{Bro,Cal}.
In general in the magnetic field operators of crystal translation do not commute \cite{Bro,Cal}.
An effective magnetic band Hamiltonian can be introduced only when the magnetic flux through any triangle
formed by two successive translations is a rational number, when expressed in terms of flux quanta \cite{Bro,Cal}.
These conditions can be satisfied only for some fields $H$ called rational by defining a superlattice structure
with a restored vector translation symmetry but a changed point group symmetry.
Superlattice calculations are more demanding because the number of inequivalent atoms in the unit cell is several times larger
and the first Brillouin zone is less symmetric. For an arbitrary magnetic field $H$ the problem can be
sorted out by approximating it with some close rational magnetic field $H' \approx H$.

Second, although the magnetic coupling is much weaker than the electric interactions including
the mean field, it is nevertheless responsible for a complete reconstruction of the energy spectrum.
Strictly speaking, we have to calculate new energy values and find new state functions explicitly depending on the value of $H$
and populate them accordingly starting from a state with lowest energy and finishing with states of highest energy ($E'_F$).
In this formulation the task has not been solved.
Nevertheless, in the past many bright ideas and nontrivial theoretical studies have been invested in this problem \cite{Rot,Blo,Wan,Sam,Gla,But}.
For example, one can start with a nearly free electron gas model assuming that the crystal periodic potential is weak \cite{Sam,Gla}.
The other, more strict approach requires calculated electron bands $E_n(\vec{k})$ in the absence of the magnetic field.
The full magnetic Hamiltonian $\hat{H}(\vec{\Pi})$ is an operator depending on $\vec{\Pi} = \vec{p} + e \vec{A}/c$,
whose components do not commute.
The operator $\hat{H}$ is a matrix function of $\vec{\Pi}$ with matrix elements depending on the magnetic field.
When written in diagonal form it can be sufficient to approximate its diagonal elements
by the effective Hamiltonian $E_n(\vec{\Pi})$ obtained by replacing $\hbar \vec{k}$ by the operator $\vec{\Pi}$
in the energy band function $E_n(\vec{k})$.
Pursuing this approach one can work out the explicit form of the effective Hamiltonian which does not couple different bands
and gives the new energy levels.
Although this approximation is considered good for isolated (non-degenerate) electron bands $E_n$,
the procedure is not rigorous.
In particular, in its derivation some terms of the order of $(\hbar \omega / E_F)^2 \sim H^2$ are omitted \cite{Zil}.

In the effective Hamiltonian treatment, the general effect of the magnetic field on electron bands \cite{Mis} is two-fold:
1) a gradual transformation
of band parameters, 2) breaking up into a series of discrete states. The latter effect is apparently
a manifestation of emerging Landau levels.
In Ref.\ \cite{Mis} the method reached the stage where in principle all necessary expressions have been
derived for the calculation of $\chi_L$ within a pseudopotential (orthogonalized plane wave)
basis set. The core formalism however has turned out to be complicated again and the authors use smallness
of pseudopotential to obtain estimates of $\chi_L$ for real metals.

Diamagnetic response in a metal can be obtained by methods of quantum field theory \cite{Heb2,Hol,Tsve}.
In Ref.~\cite{Heb2} Fourier components of the induced current are calculated through the evolution of
the single particle density matrix.
The orbital susceptibility $\chi_{dia}(q,\omega)$
is found as a prefactor for $q^2$ ($q$ is the wave vector, $\omega$ is the field frequency).
The Landau susceptibility $\chi_{L}$ of free electrons is restored at $\omega = 0$
and $q \rightarrow 0$ as a result of cancellation of two terms which diverge as $1/q^2$.
In Ref.\ \cite{Tsve} and \cite{Hol} a model of conduction electrons interacting with an electromagnetic field
is considered. Of particular interest is the electron interaction via
current-current potential arising from the exchange of transverse photons.
When Dyson equations for complete two-point correlation functions are solved,
the Landau susceptibility appears in the corresponding polarization operator \cite{Tsve}.
The diamagnetic response comes from the states at the
Fermi surface and in general contain terms with the derivatives of the
density of states. For arbitrary dispersion however, such terms cannot be
expressed as derivatives of the fermionic energy.

There is another approach to the problem based on the semiclassical treatment of the problem \cite{Ons,Lif,Sho,AM,semi}.
For example, it has proven to be highly effective in describing
the de Haas - van Alphen oscillatory part of the magnetic susceptibility.
Onsager \cite{Ons} and Lifshitz \cite{Lif} based on the semi-classical description of the
movement of an electron in a magnetic field, showed that the change in $1/H$ is determined by extremal cross-sections of the Fermi surface
in a plane normal to the magnetic field.
A good historical and theoretical review of the effect is given in the book of Shoenberg \cite{Sho}.
Although the de Haas - van Alphen effect can be understood within the quantum theory \cite{Lan0,Hol},
the full quantum consideration is not extended to real crystals and in practice
the description of the de Haas - van Alphen oscillations relies on the semiclassical theory \cite{Sho}.

From the semiclassical equations it follows that the component of $\vec{k}$ (in the reciprocal space)
parallel to $\vec{H}$
and the electron energy $E(\vec{k})$ are both constants of the electron motion.
Therefore, in $k-$space electrons move along curves defined by the intersection of isoenergetic surfaces ($E(\vec{k}) = E_0$) with planes
perpendicular to the magnetic field, i.e. the scalar potential associated with the electron mean field in solid is unchanged.
This is a reasonable assumption especially in the limit of small magnetic fields $\hbar \omega / E_F \rightarrow 0$,
which is the case of our consideration.

Electron movement along a closed trajectory (orbit) is a subject of the quantization conditions, Eqs.\ (\ref{l1}), (\ref{l1b}) below.
Although these quantization conditions are familiar from the semiclassical approach, they also can be derived from
an equation-of-motion method \cite{semi,Rot2}.
In this paper we will use this semiclassical approach to obtain a very simple expression for the Landau diamagnetic susceptibility
of real metals at zero temperature (although the method can be extended to the case of a finite temperature).
The elementary unit in our approach is a magnetic tube of finite width and length,
containing a certain Landau level inside it.
The occupied tubes whose electron states are completely filled are diamagnetically inert.
Only partially occupied tubes located in the neighborhood of the Fermi surface contribute to the diamagnetic effect.

Two important consequences follow from our approach. First, at zero temperature the Landau diamagnetic susceptibility is
caused by electron states near the Fermi energy. It is worth noting that the oscillatory part of the diamagnetic
susceptibility giving rise to the de Haas - van Alphen effect has been related to the extremal orbits of the Fermi surface
for many years. Here we accomplish the relation by ascribing also the steady diamagnetism to the Fermi surface states.
The leading role of the Fermi surface in diamagnetism is in accord with the other properties of conduction electrons
(for example, the Pauli paramagnetism) which
also depend on the peculiarities of electron states near the Fermi energy \cite{Zim}.
Second, the presented method can be applied to {\it ab initio} calculations of the steady diamagnetic susceptibility of real metals.
In this paper we consider the simplest case, which is the case of monovalent alkali metals.
We hope that in principle based on our simple expression (\ref{e20}) below, the Landau susceptibility
can be added to a list of physics quantities available from first principles calculations.

We should also mention the important question of the Fermi energy ($E_F$) change in the applied magnetic field.
This is a very weak effect, considered first in detail by Kaganov {\it et al}. \cite{Kag}.
Some consequences of the phenomenon including weak oscillations of the density of states at the Fermi level
are further discussed by Shoenberg in Ref.\ \cite{Sho}.
Below we will unveil a mechanism of this effect for the free electron gas case.  It is caused by peculiarities of electron population
in the equatorial region of the Fermi sphere. Depending on the applied magnetic field there can be a small inflow or outflow of electrons
from the equatorial region to other states of the Fermi surface.

The paper is organized as follows.
In Sec.\ \ref{sec:feg} we demonstrate the method for the free electron gas.
There, we introduce a central object for our method -- a magnetic tube
and consider its properties. The most important one is that the tube whose electron states
are completely occupied does not contribute to magnetic response. Then we select
diamagnetically active tubes near the Fermi surface and calculate their steady diamagnetic susceptibility $\chi_{L}$.
The oscillatory (de Haas-van Alphen) contribution ($\chi_{dHvA}$) is considered in section \ref{sec:dHvA}.
Here the analytical calculations become more involved although the general physical picture remains clear and transparent.
The description of the steady diamagnetism of real metals, Sec.\ \ref{sec:realmet}, follows the same lines.
As in the case of the Fermi gas, we first consider magnetic tubes of real systems and then
calculate their steady diamagnetic response.
The application of the method for calculations of $\chi_L$ for various magnetic field directions
is given in Sec.\ \ref{sec:real}.
Finally, in Sec.\ \ref{sec:con} we summarize main conclusions.

\section{Free electron gas}
\label{sec:feg}

\subsection{Magnetic tubes in $\vec{k}$-space}
\label{ssec:gast}

In an external magnetic field $\vec{H}$ directing along the $z$-axis,
the energy of the free electron is given by \cite{Lan0}
\begin{eqnarray}
     E = \hbar \omega \left( n+\frac{1}{2} \right) + \frac{\hbar^2 k_z^2}{2 m} ,
\label{g0}
\end{eqnarray}
where $n$ is integer (numbering the Landau levels), $k_z$ is the $z-$component of the wave vector $\vec{k}$,
and the cyclotron frequency
\begin{eqnarray}
     \omega = \frac{eH}{mc} .
\label{g4}
\end{eqnarray}
Here $m$ and $e$ are the electron mass and charge; $c$ is the speed of light.
In correspondence with Eq.\ (\ref{g0}) the energy of electron is presented by two contributions,
the contribution $E_{\perp}$ from the movement in the plane, perpendicular to $\vec{H}$ (i.e. in the plane $(k_x,k_y)$)
and the contribution $E_z = \hbar ^2 k_z^2 /2m$ from the movement parallel to $\vec{H}$ (i.e. along the $z$-axis).
In the following we consider only the component  $E_{\perp}$, because the parallel component $E_{\parallel}=E_z$
is unchanged in the magnetic field.

Although Eq.\ (\ref{g0}) is obtained within the fully quantum treatment,
below we will follow the wide used semiclassical representation of electron orbits in
the real and momentum ($\hbar \vec{k}$) space \cite{Ons,Lif,AM,semi}, which gives exactly
the same energy spectrum.
The electron movement in the magnetic field in the $(x,y)$ plane along the radial direction $q=r$ or along
the $q=x$ or $y$-axis (depending on the choice of the vector potential gauge)
is described by a quantized orbit,
\begin{eqnarray}
     \oint p_q\, d q = 2\pi \hbar\, \left( n +\gamma \right),
\label{l1}
\end{eqnarray}
which corresponds to the $n$-th Landau level with the factor $\gamma$.
In Wentzel, Kramers, Brillouin' (WKB) method for non-singular integrable orbits $\gamma = \mu/2$ \cite{semi},
where $\mu$ is the Maslov index. In our case $\mu=2$ and $\gamma=1/2$ \cite{semi}.
In the $(k_x, k_y)$ momentum plane the electron orbit defined by Eq.\ (\ref{l1}) is visualized as a circle
whose area $A_n$ is given by
\begin{eqnarray}
     A_n = \frac{2\pi eH}{c \hbar} \left( n +\frac{1}{2} \right) ,
\label{l1b}
\end{eqnarray}
and its energy is
\begin{eqnarray}
     E_{\perp,n} = \hbar \omega \left( n+\frac{1}{2} \right) ,
\label{gg3}
\end{eqnarray}

\begin{figure}[ht]\center
\begin{tabular}{l r}
\includegraphics[width=30mm]{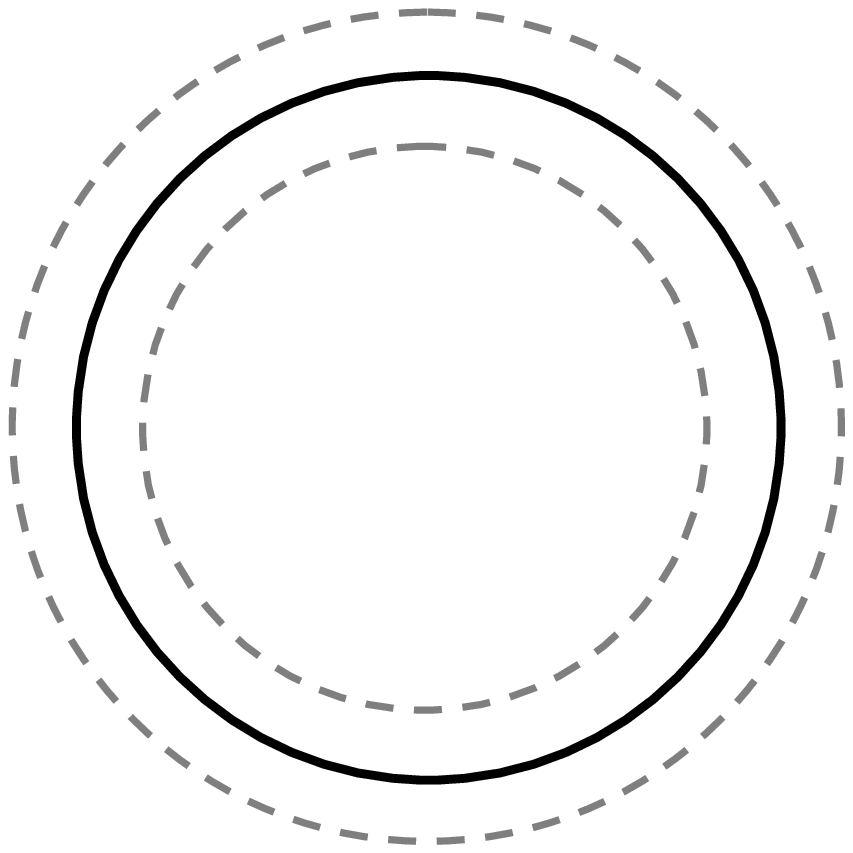}
&
\includegraphics[width=30mm]{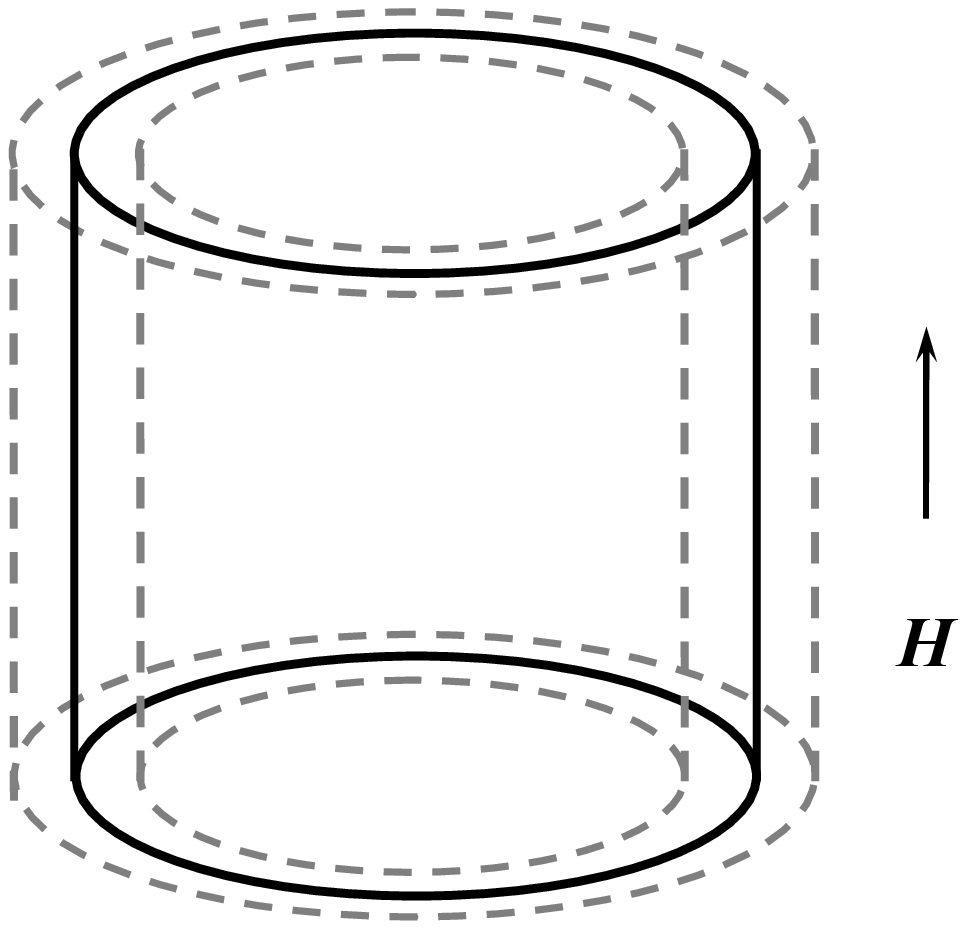}
\end{tabular}
%
\caption{Magnetic tube and the corresponding Landau level.
On the left: the $(k_x,k_y)$ tube cross-section and the $n$-th Landau orbit (the solid circle with the in-plane energy $E_n$).
The dashed circles correspond to the auxiliary orbits with energies $E_n^{aux}$ and $E_{n+1}^{aux}$.
On the right: the tube in the $\vec{k}$-space.
Without magnetic field electron states are distributed throughout the whole tube, in the presence --
only on the Landau orbit in the middle.
}
 \label{fig0_1}
\end{figure}

It is well known that the average density of electron states in the $\vec{k}$-space remains the same
as without magnetic field.
To understand better the reconstruction of the electron structure in the magnetic field $H$,
we select in the $\vec{k}$-space a tube, whose number of electron states and the energy of all states
do not change in the presence of $H$, Fig.\ \ref{fig0_1}.
For that we consider auxiliary electron orbits of the area
\begin{eqnarray}
       A_n^{aux} = \frac{2\pi e H}{ c \hbar }\, n
\label{l5a}
\end{eqnarray}
with energies
\begin{eqnarray}
     E_n^{aux} = \hbar \omega n .
\label{l5}
\end{eqnarray}
Note the the $n$-th Landau orbit defined by Eqs.\ (\ref{l1b}), (\ref{gg3}) is situated between the auxiliary orbits $n$ and $n+1$,
Fig.~\ref{fig0_1} left panel,
and its energy $E_n$ lies between $E_n^{aux}$ and $E_{n+1}^{aux}$.
Below we show that the number of electron states with energies $E_n^{aux} \leq E \leq E_{n+1}^{aux}$ without field
equals the number of electron states condensing on the $n$-th Landau level in the presence of the field.
The same holds for their total energies.

For that we calculate the density of electron states ${\cal N}_{\perp}$ in the $(k_x,k_y)$-plane,
\begin{eqnarray}
     {\cal N}_{\perp} = \frac{d N_{\perp} }{d E_{\perp} } = \frac{2m}{\hbar^2} \frac{L_x L_y}{2\pi} .
\label{lg9}
\end{eqnarray}
and notice that ${\cal N}_{\perp}$ is independent of the energy $E_{\perp}$.
(Here $L_x$, $L_y$ and $L_z$ are distances of the free electron gas box in $x$, $y$ and $z$ directions, respectively.)
Using (\ref{lg9}),
we find the number of electron states in the $n$-th tube without field, i.e.
in the energy range $E_n^{aux} \leq E \leq E_{n+1}^{aux}$,
\begin{subequations}
\begin{eqnarray}
     \triangle N_n(H=0) = \int_{E_n^{aux}}^{E_{n+1}^{aux}} {\cal N}_{\perp}\, dE_{\perp} = {\cal N}_{\perp}\, \hbar \omega = 2 N_p . \quad
\label{l10a}
\end{eqnarray}
Here $N_p$ is the spatial degeneracy of the Landau levels (without spin polarization),
\begin{eqnarray}
     N_p = \frac{L_y}{2\pi}\, \frac{m\omega}{\hbar} L_x = \frac{L_x L_y}{2\pi} \frac{eH}{c \hbar}.
\label{l8}
\end{eqnarray}
Calculating the total energy of these states without field,
\begin{eqnarray}
   & &{\cal E}_n(H=0) = \int_{E_n^{aux}}^{E_{n+1}^{aux}} {\cal N}_{\perp} E_{\perp} dE_{\perp}   \nonumber \\
   & & =  \frac{1}{2}\left[ (E_{n+1}^{aux})^2 - (E_n^{aux})^2 \right] {\cal N}_{\perp}
   = E_{\perp,n} \triangle N_n , \quad
\label{l10b}
\end{eqnarray}
\end{subequations}
we find that it coincides with the energy of all electron tube states condensed on the $n$-th Landau level in the presence of the field.
Thus, we have proven that
\begin{subequations}
\begin{eqnarray}
  & &   \triangle N_n(H=0) = \triangle N_n( H \neq 0) = 2 N_p , \label{g11a} \\
  & &   {\cal E}_n(H=0) = {\cal E}_n(H \neq 0) = \hbar \omega \left( n+\frac{1}{2} \right) 2 N_p . \quad \quad \label{g11b}
\end{eqnarray}
\end{subequations}

%
%
\begin{figure}
\resizebox{0.4\textwidth}{!} {
\includegraphics{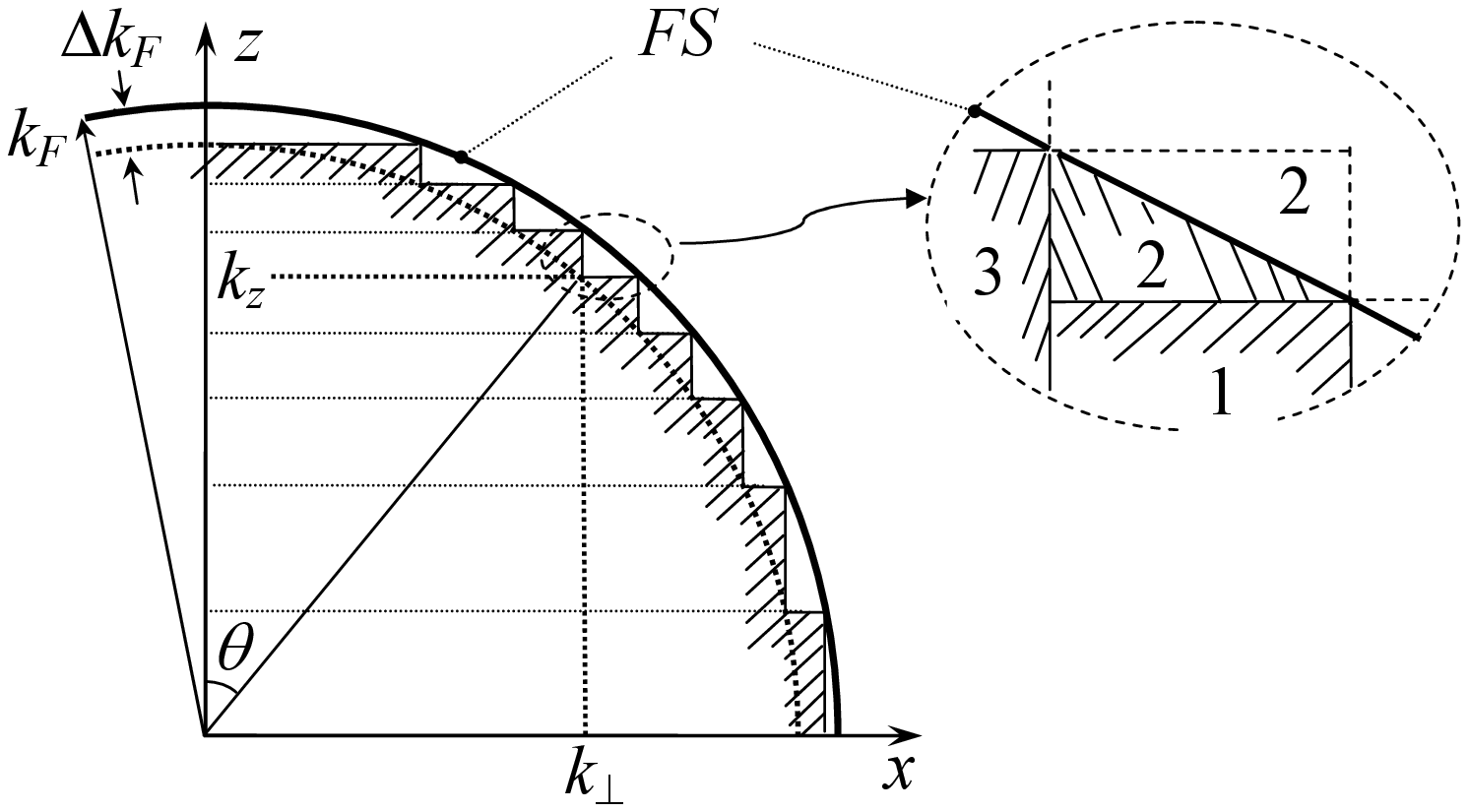}   }

\vspace{2mm}
\caption{
Magnetic tubes and the Fermi surface ($FS$), the $(k_x,k_z)$-cross-section in the $\vec{k}$-space.
The tubes inside the Fermi sphere shown as dashed area on the left panel, are completely filled
and diamagnetically inert.
Inset: 1 - completely occupied tube $n$, 2 - partially occupied tube $n$, 3 - completely occupied tube $(n-1)$.
} \label{fig0_2}
\end{figure}

So far our consideration has been limited by the $(k_x,k_y)$ plane.
However, since Eqs.\ (\ref{g11a}) and (\ref{g11b}) hold for any $k_z-$component, they are fulfilled for the whole tube $n$, Fig.\ \ref{fig0_1}.
In other words, if without field, a tube contains electron states which satisfy the inequality $E_n^{aux} \leq E_{\perp} \leq E_{n+1}^{aux}$
for all $k_z$ in the range $k_z^{(1)} \leq k_z \leq k_z^{(2)}$, then in the presence of the magnetic field all these states condense on
the $n$-th Landau level, that is, $E_{\perp}(k_z) = E_{\perp,n}(k_z)$ throughout the $k_z$ region, Fig.\ \ref{fig0_1}.
Furthermore, the total electron energy of the tube remains unchanged, which bears some resemblance to the Bohr - van Leeuwen theorem
in classical physics.
The upper $k_z^{(2)}$ and lower $k_z^{(1)}$ boundary of the tube can be taken arbitrary.
For practical reason for each tube $n$ we define the $k_z$-boundaries by its intersection with the Fermi surface, Fig.\ \ref{fig0_2}.
We then obtain two $n$-th tubes: the first tube (denoted by $1$ in the inset of Fig.\ \ref{fig0_2}) lies entirely inside the Fermi surface and
being completely occupied does not exhibit diamagnetism.
The second tube (denoted by $2$ in the inset of Fig.\ \ref{fig0_2}) being only partially filled below the Fermi surface,
results in a diamagnetic response.
We consider this effect in the following sections.
%
%

It is also worth noting that the $\vec{k}$-space partitioning depends on the value of the magnetic field, since $\omega \sim H$,
and the tube boundaries are defined by $\omega$, Eq.\ (\ref{l5}).

\subsection{Diamagnetically active electron states}
\label{ssec:gass}

Consider the Fermi surface and define necessary magnetic tubes parallel to the $z$-axis (in the direction of the magnetic field $H$), Fig.\ \ref{fig0_2},
as discussed in Sec.\ \ref{ssec:gast}.
Boundary conditions defined by in-plane circular orbits, Eq.\ (\ref{l5a}), specify a set of concentric cylindrical surfaces,
which intersect the Fermi surface in circles perpendicular to the $z$-axis.
We then draw the planes of the circles and use them to construct a set of tubes, limited by the planes and the cylindrical surfaces, which lie
inside the Fermi sphere.
The $(k_x, k_z)$ cross-section of these tubes is schematically shown in Fig.\ \ref{fig0_2}.
The fully occupied tubes are shown as dashed area.
The electron states of the completely filled tubes do not change their energy in a magnetic field.
Therefore, the whole effect is due to the states lying in the partially occupied tubes.
Their cross-sections in the $(k_x, k_z)$-plane look like a chain of triangles, Fig.\ \ref{fig0_2}.

\begin{figure}
\resizebox{0.25\textwidth}{!} {
\includegraphics{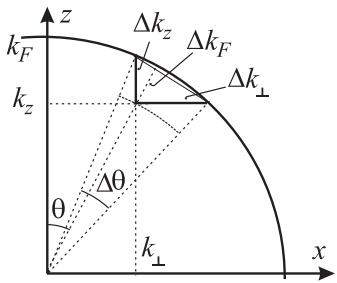}   }

\caption{
A typical $(k_x,k_z)$-cross-section of a partially occupied tube near the Fermi surface.
The triangle size is greatly exaggerated, since $\triangle k_F = m \omega/\hbar k_F \ll k_F$.
} \label{fig0_3}
\end{figure}

Consider a typical partially occupied tube, whose triangle cross-section in the $(k_x, k_z)$-plane is shown in Fig.\ \ref{fig0_3}.
We denote two legs of the triangle by $\triangle k_{\perp}$ and $\triangle k_z$.
Taking into account that the area of the $(k_x, k_y)$ cross-section of the $n$-th tube is $\triangle A =  A_{n+1}^{aux} - A_n^{aux} = 2\pi e H/ c \hbar$,
and that $\hbar \omega \ll E_F$, we obtain
\begin{eqnarray}
 \triangle k_{\perp} =  \frac{\triangle A}{2\pi k_{\perp}} =  \frac{m \omega}{\hbar k_F} \frac{1}{\sin \Theta} .
 \label{f3}
\end{eqnarray}
(Here $\Theta$ is the polar angle, Figs.\ \ref{fig0_2}, \ref{fig0_3}.)
Therefore, the narrow surface region of the partially occupied tube is defined by
the wave vector quantity $\triangle k_F$ shown in Figs.\ \ref{fig0_2} and \ref{fig0_3},
\begin{eqnarray}
 \triangle k_F  = \triangle k_{\perp} \sin \Theta = \frac{m \omega}{\hbar k_F}  .
 \label{f4}
\end{eqnarray}
It is remarkable that $\triangle k_F$ is independent of $\Theta$.
Therefore, the radius $k_F - \triangle k_F$ determines an auxiliary internal sphere in the $\vec{k}$-space,
which can be used for
drawing the step-wise line shown in Figs.\ \ref{fig0_2} and \ref{fig0_3},
separating the fully occupied tubes from the partially occupied ones.

For the angle $\triangle \Theta$ shown in Fig.\ \ref{fig0_3}, we obtain
\begin{eqnarray}
   \triangle \Theta = \frac{\triangle k_F}{k_F} \frac{1}{\cos \Theta \sin \Theta} . \label{f10b}
\end{eqnarray}
Notice, that Eq.\ (\ref{f3}) has a singularity at $\Theta = 0$, and Eq.\ (\ref{f10b}) at $\Theta = 0$ and $\pi/2$.
Therefore, the polar and equatorial region of the Fermi sphere should be considered more attentively,
see Sec.\ \ref{sec:dHvA} below.

Now we find the number of active electron states in the partially occupied tubes,
\begin{eqnarray}
  N = \sum_{n=1}^M \triangle N_n ,
 \label{g11}
\end{eqnarray}
where $\triangle N_n \equiv \triangle N(k_z,\; k_z + \triangle k_z)$ is the number of the electron states
in the $n$-th partially filled tube, whose $k_z-$component lies between $k_{z,n} \equiv k_z$ and $k_{z,n+1} \equiv k_z + \triangle k_z$.
Using the infinitesimal property of the $(k_x, k_y)$ cross-section we find
\begin{eqnarray}
   \triangle N = 2v \triangle V_k = 2\pi v \frac{eH}{\hbar c}  \triangle k_z , \label{g10a}
\end{eqnarray}
where $\triangle V_k$ is the volume of the partially occupied tube in $k-$space and $v = V/(2\pi)^3$
(In the general case $\triangle N$ is calculated in Appendix \ref{subsec:st}.)

Notice that from Eq.\ (\ref{g10a}) it follows that $\triangle N / \triangle k_z = const$.
Furthermore, $\triangle N / \triangle k_z$ is independent of $m$, which will be fully appreciated in the general case
considered below in Sec.\ \ref{sec:pert}.
Since for usual magnetic fields $\triangle k_F,\; \triangle k_z \ll k_F$,
in Eq.\ (\ref{g11}) we can substitute the summation with the integration,
\begin{eqnarray}
  N = \int d N = \int_{k_z^{min}}^{k_z^{max}} \left( \frac{\triangle N}{ \triangle k_z } \right) d k_z  .
 \label{g12a}
\end{eqnarray}
For the free electron gas $k_z^{min} = -k_F$, $k_z^{max} = k_F$.
Using (\ref{g10a}), (\ref{g12a}) we arrive at
\begin{eqnarray}
  N = 2\pi v \frac{e H}{\hbar c}\, \ell_z  ,
 \label{f15}
\end{eqnarray}
where $\ell_z \equiv k_z^{max} - k_z^{min} = 2k_F$ is defined exclusively by the projection of the Fermi surface
along the direction of the applied magnetic field.
Therefore, $N \sim H$.
Since the perturbation energy for each electron state can be estimated with $\hbar^2 k_F \triangle k_F/m =\hbar \omega \sim H$,
the total energy change in the magnetic field $\sim H^2$, which leads to the constant magnetic susceptibility $\chi$.
(The rigorous computation of $\chi$ is given in the next section.)
%
%

\subsection{Steady diamagnetic susceptibility}
\label{ssec:gasd}

%
\begin{figure}
\resizebox{0.4\textwidth}{!} {
\includegraphics{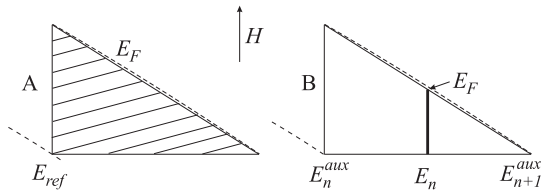}   }

\vspace{2mm}
\caption{
The $(k_x,k_z)$-cross-section of a partially occupied tube near the Fermi level.
On the left ($A$): dashed area -- occupied electron states without magnetic field, $\vec{H}=0$.
On the right ($B$): bold vertical line -- the occupied electron states (1D gas) in the magnetic field $\vec{H} \neq 0$.
$E_F$ is the Fermi energy of the one-dimensional electron gas (along the $z$-axis), whose transverse energy $E_n$
is determined by the $n$-th Landau energy.
} \label{fig0_4}
\end{figure}
%

A remarkable property of partially occupied tubes near the Fermi surface is that
application of a magnetic field does not lead to electron transitions between different tubes
(with the exception of a small number of electrons in the equatorial region).
Therefore, upon applying the field, there is a redistribution of electron states only within each
partially filled tube.

To demonstrate this, we consider in detail the transformation of electron states in the tube $n$
when the magnetic field is switched on.
(Without field the number of electron states $\triangle N$ is given by Eq.\ (\ref{g10a}).)
The occupation of electron states for two cases ($H = 0$ and $H \neq 0$) is shown schematically in Fig.\ \ref{fig0_4}.
When $H \neq 0$, all electrons of the tube are on the $n$-th Landau level with the transverse energy $E_n$,
and occupy the lowest $k_z$-states, as shown in Fig.\ \ref{fig0_4}.
If all electrons remain in the tube, then the highest energy level with the wave vector
$\delta k_z$ [with to $k_z$ of the $n-$th tube, Fig.\ \ref{fig0_3}] is found from the following relation:
\begin{eqnarray}
   2 \delta k_z \, N_p\, l_z = \triangle N .
 \label{e12}
\end{eqnarray}
We recall that $N_p$ is the in-plane (or transverse) folding of the $n$-th Landau level,
while $l_z=L_z/2\pi$ is the density of electron states along $k_z$.
Substituting in Eq.\ (\ref{e12}) Eq.\ (\ref{l8}) for $N_p$
and Eq.\ (\ref{g10a}) for $\triangle N$, we obtain
\begin{eqnarray}
   \delta k_z = \frac{1}{2} \triangle k_z .
 \label{eg12a}
\end{eqnarray}
(The result for the infinitely small triangle cross-section can be foreseen from the geometrical reasons.)

Eq.\ (\ref{eg12a}) leads to an important consequence. The energy of the highest occupied electron level coincides with
$E_F$ and the wave vector $k_F$ lies on the Fermi surface even in the applied magnetic field $H \neq 0$.
Since the conclusion holds for all partially occupied tubes (with the exception of few equatorial tubes),
the highest energy of the occupied electron states -- $E_F$ -- is conserved as the Fermi energy for all tubes and there are no electron transitions
between tubes. (The exceptional case of the equatorial region is considered later in Sec.\ \ref{sec:dHvA}.)
Therefore, in the following we can calculate the energy change for each tube separately.

Keeping in mind that in the magnetic field there are two energy contributions:
$E_{\perp}$ in the $(k_x,k_y)$ plane and $E_z$ at $k_z$,
we obtain for the energy change of the partially filled tube $n$,
\begin{eqnarray}
  \triangle E
  = \triangle E_{\perp}^{H \neq 0} + \triangle E_z^{H \neq 0} - \triangle E^{H = 0} . \quad
 \label{eg2n}
\end{eqnarray}
The notation $\triangle$ on the right hand side is an indication that the corresponding energy refers to the partially filled tube $n$,
whose $k_z$ values are in the range from $k_{z,n}$ to $k_{z,n+1}=k_{z,n} + \triangle k_z$, Fig.\ \ref{fig0_2}, \ref{fig0_3}.
The quantities $\triangle E_{\perp}^{H \neq 0}$, $\triangle E_z^{H \neq 0}$, and $\triangle E^{H=0}$,
therefore refer to energy components of the tube with ($H \neq 0$) and without ($H = 0$) magnetic field.

The detailed simple calculations of all components are performed in Appendix \ref{subsec:en}
(with respect to the energy $E_{ref}$, Fig.\ \ref{fig0_4}).
As a result, we get
\begin{eqnarray}
  \frac{\triangle E^{H=0}}{\triangle N} =
   \frac{2}{3} \frac{\hbar^2}{m} k_F \triangle k_F = \frac{2}{3} \hbar \omega
 \label{eg10}
\end{eqnarray}
for the energy components without magnetic field and
\begin{subequations}
\begin{eqnarray}
  \frac{\triangle E_{\perp}^{H \neq 0}}{\triangle N} = \frac{1}{2} \hbar \omega ,  \label{eg11a} \\
  \frac{\triangle E_z^{H \neq 0}}{\triangle N} = \frac{1}{4} \hbar \omega          \label{eg11b}
\end{eqnarray}
\end{subequations}
in the applied magnetic field.
In fact, taking into account that $\triangle N$ stands for the number of electrons,
the right hand sides of Eqs.\ (\ref{eg10}), (\ref{eg11a}) and (\ref{eg11b})
represent average energy values independent on the tube under consideration.
The substitutions of (\ref{eg10}), (\ref{eg11a}) and (\ref{eg11b}) in (\ref{eg2n}) yields
\begin{eqnarray}
  \triangle E = \frac{1}{12} \hbar \omega\, \triangle N > 0 .
 \label{eg18}
\end{eqnarray}

Note that Eq.\ (\ref{eg18}) refers to any partially filled tube.
Therefore, making the summation over all tubes and using Eq.\ (\ref{f15}) with $\ell_z = 2k_F$, we find
\begin{eqnarray}
  E = \frac{1}{12} \hbar \omega\, N = \frac{1}{3}\, \pi v\, k_F \, \frac{e^2 H^2}{m c^2} .
 \label{eg19}
\end{eqnarray}
For the magnetic susceptibility $\chi$
we finally have:
\begin{eqnarray}
  \chi = -\frac{d^2E(H)}{dH^2} = -\frac{2}{3}\, \pi v \, k_F \frac{e^2}{mc^2} =-\frac{e^2 k_F V}{12 \pi^2 mc^2} . \nonumber \\
 \label{g20}
\end{eqnarray}
This is the celebrated expression, obtained by Landau for the diamagnetic susceptibility of the free electron gas.

\subsection{Oscillatory (equatorial) diamagnetic susceptibility}
\label{sec:dHvA}

Earlier (Sec.\ \ref{sec:Lan}) we have obtained the diamagnetic effect based of calculations of the energy of active electrons
in the partially occupied tube of general form.
Deviations from the general situation are possible for boundary cases, which are the polar region ($\Theta=0$) with the Landau level $n=0$,
and the equatorial region ($\Theta=\pi/2$).
In Appendix \ref{subsec:polar} we analyse the polar region and show that it complies with the general case.
In the equatorial region however the situation is very different.
%
\begin{figure}
\resizebox{0.2\textwidth}{!} {
\includegraphics{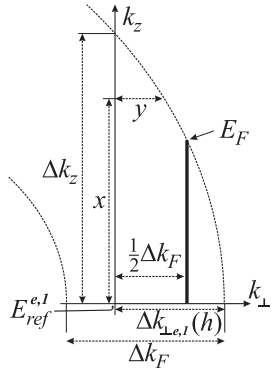}   }

\vspace{2mm}
\caption{
First equatorial tube.
Landau states (shown by the bold line) are occupied if $\triangle k_{\perp,e,1}=h > \triangle k_F/2$ (or $r > 1/2$),
and empty if $h < \triangle k_F/2$ ($r < 1/2$).
} \label{fig0_5}
\end{figure}
%

The problem is that the step-wise line shown in Fig.\ \ref{fig0_2} can terminate at the equatorial point with $\Theta=\pi/2$
in any place with $k_{\perp,e}$ lying in the interval $k_F - \triangle k_F \le k_{\perp,e} \le k_F$,
and the equatorial point does not necessarily lie on the internal sphere of the radius $k_F - \triangle k_F$,
which is the case for all other tubes, Fig.\ \ref{fig0_5}.
This equatorial tube is truncated because its upper energy boundary  $E_{n+1}^{aux}$, defined by (\ref{l5}),
in general lies outside the $(k_x,k_y)$ equatorial cross-section of the Fermi sphere and cannot be reached.
We define this irregular tube with $k_{\perp} \ge k_{\perp,e}$ as the first equatorial tube.
Notice that when $k_{\perp,e} \rightarrow k_F$, the area of the $(k_x,k_z)$ cross-section of this tube approaches zero.
In that case one has to resort to the preceding tube (that is, with $k_{\perp} < k_{\perp,e}$),
which also makes an irregular contribution to energy.
We define it as the second equatorial tube.
The other tubes essentially follow the general dependencies considered in Sec.\ \ref{sec:Lan}.

For the first equatorial tube we define the quantity $\triangle k_{\perp,e,1} = k_F - k_{\perp,e}$, Fig.\ \ref{fig0_5},
for which we shall also use a short notation $h = \triangle k_{\perp,e,1}$.
The subscripts $e,1$ and $e,2$ refer to the
first and second equatorial tube, respectively.
As discussed above,  $h$ ranges from 0 to $\triangle k_F$.
Consider the important dimensionless parameter
\begin{eqnarray}
  r = \frac{h}{\triangle k_F} .
 \label{q2}
\end{eqnarray}
Clearly, $0 < r < 1$.
Note, that by varying $H$ we change the structure of all magnetic tubes, and, consecutively
the parameter $r$, defined by the geometry of the last tube.
Therefore, $r$ implicitly depends on  $H$. It can be shown
that in a first approximation $r$ is proportional to $1/H$.

For $\triangle k_{z,e,1}$ we obtain
\begin{eqnarray}
  \triangle k_{z,e,1} = \sqrt{2 k_F h} = \sqrt{2 r \frac{eH}{\hbar c}} .
 \label{q1}
\end{eqnarray}
Calculating the number of states in the first equatorial tube without magnetic field, we find
\begin{eqnarray}
  \triangle N_{e,1}^{H=0} = \frac{8 \pi}{3} rv \frac{eH}{\hbar c}\, \triangle k_{z,e,1}.
 \label{qq5}
\end{eqnarray}
Based on the analysis of the Cornu spiral sum, Pippard estimated that the relative weight
of the extremal region should be $\triangle N_e / N \sim \sqrt{H}$ (Eq.\ (33) of Ref.\ \cite{Pip}).
This conclusion is in agreement with Eq.\ (\ref{qq5}) since
$\triangle N_{e,1}^{H=0}/N \sim \triangle k_{z,e,1} \sim \sqrt{\omega} \sim \sqrt{H}$.

Notice that already in obtaining $\triangle N_{e,1}^{H=0}$ we have a deviation from the general case, Eq. (\ref{qq5}), and
\begin{eqnarray}
  \frac{\triangle N_{e,1}^{H=0}}{\triangle \Theta_{e,1}} = \frac{4r}{3} \frac{\triangle N}{\triangle \Theta} .
 \label{qq7}
\end{eqnarray}
Here $\triangle \Theta_{e,1}$ defines the angular size of the first equatorial triangle in the $(k_x,k_z)$-cross-section.
Deviations are also present for the transverse and $z-$energy contributions.

Now we consider the situation in the magnetic field $H \neq 0$.
We start as in Sec.\ \ref{sec:Lan} with finding the wave vector $\delta k_z$ of the highest occupied
electron state along the $z$-axis under assumption that all electrons belonging to the first equatorial tube
do not leave it.
By means of Eq.\ (\ref{e12}) we get
\begin{eqnarray}
  \delta k_{z,e,1} = \frac{2}{3}\, r \, \triangle k_{z,e,1} .
 \label{q9}
\end{eqnarray}
Now however the energy of the highest occupied state in general differs from $E_F$, and
therefore from the energy of the highest occupied states in other tubes, Eq.\ (\ref{eg12a}).
Below we consider the situation for two different cases: $0 \le r < 1/2$ (case $a$) and $1/2 \le r < 1$ (case $b$).

In case $a$ the energy of the Landau level of the first equatorial tube $E_n - E_{ref}^{e,1} = \hbar \omega/2$, Fig.\ \ref{fig0_5},
is higher than $E_F$ even at $k_z = 0$.
Therefore, all electrons from this tube move to other tubes where they occupy free states above $E_F$.
As a result, a small rise in $E_F$ should occur, but since $\triangle N_{e,1} \ll N$, it is of the order of
$\hbar \omega \, \triangle N_{e,1} / N \ll \hbar \omega$.
Since $E_F - E_{ref}^{e,1} = r \hbar \omega$, the energy of the promoted electrons is
$\triangle E^{H \neq 0,a}/\triangle N_{e,1}^{H=0} = r \, \hbar \omega$
(with respect to $E_{ref}^{e,1}$).

\begin{figure}
\resizebox{0.4\textwidth}{!} {
\includegraphics{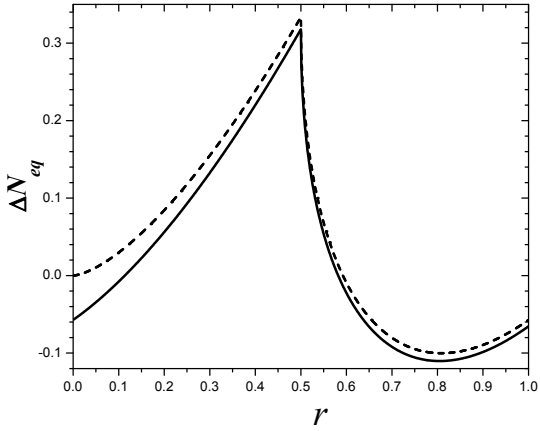}   }

\caption{
The number of electrons (in units of $8 \pi v (eH / \hbar c)^{3/2}$) promoted from the equatorial range to other tubes,
$\triangle N_{eq}=\triangle N_{e}^{H = 0} - \triangle N_{e}^{H \neq 0}$, expressed
in terms of the dimensionless parameter $r \sim 1/H$.
Negative values imply that electrons move to the equatorial tube.
The dashed line stands for the contribution from the first equatorial tube, solid line -- from the first and second equatorial tubes.
The same plot (in units of $(eH / \hbar c)^{3/2} \hbar^2/ m k_F$) describes a small oscillatory dependence
of the Fermi energy (chemical potential).
} \label{fig0_6}
\end{figure}

In case $b$ the Landau level at $k_z = 0$ lies below $E_F$ and in the magnetic field it becomes partially occupied by electrons with $k_z > 0$.
The maximal $z$-wave vector $\delta k_{F}^z$ of the 1D electron state lying on the Fermi sphere can be found
by requiring its energy to be equal to $E_F$,
\begin{eqnarray}
  \delta k_{F}^z = \sqrt{2 k_F \left(h - \frac{1}{2} \triangle k_F \right) }
  = \triangle k_{z,e,1} \sqrt{\frac{r - \frac{1}{2}}{r}} .
 \label{q11}
\end{eqnarray}
The number of the occupied electron states in the tube, $\triangle N_{e,1}^{H \neq 0,b}$,
is determined by
\begin{eqnarray}
  \frac{ \triangle N_{e,1}^{H \neq 0,b} }{\triangle N_{e,1}^{H=0}} = \frac{3}{2r^{3/2}} \sqrt{r - \frac{1}{2}} .
 \label{q12}
\end{eqnarray}
The condition $\triangle N_{e,1}^{H=0} > \triangle N_{e,1}^{H \neq 0,b}$ in terms of $r$ means $1/2 \le r < \sqrt{3} \sin \pi /9$,
while $\triangle N_{e,1}^{H=0} \le \triangle N_{e,1}^{H \neq 0,b}$ results in $\sqrt{3} \sin \pi /9 \le r < 1$.
Therefore, if $1/2 \le r < 0.529$, electrons from the first equatorial tube partially move to other (regular) tubes as happens in case $a$.
For $0.529 \le r < 1$ the opposite happens, that is a small number of electrons from all regular tubes move to the equatorial tube.
The change of the number of electrons in the equatorial region is shown in Fig.\ \ref{fig0_6}.
%
%

To single out the irregular contribution explicitly, we rewrite it
in the following form,
\begin{eqnarray}
   E = E_L + \triangle E_{eq} .
 \label{q19}
\end{eqnarray}
Here $E_L$ is the diamagnetic (regular) contribution, Eq.\ (\ref{eg19}), and $\triangle E_{eq}$
stands for the irregular term from the equatorial region.
If only the first equatorial tube is accounted for, then $\triangle E_{eq} = \triangle E_{eq,1}$,
where
\begin{eqnarray}
 \triangle E_{eq,1} &=& \triangle E_{\perp,e,1}^{H \neq 0}   + \triangle E_{z,e,1}^{H \neq 0}
 - \triangle E_{e,1}^{H=0}   \nonumber \\
  & & + \triangle E_{pr,1} -\triangle E_{corr,1} .
 \label{q20}
\end{eqnarray}
Here $\triangle E_{pr,1}$ is the energy of the promoted electrons (transferred to or from regular tubes),
while $\triangle E_{corr,1}$ stands for the regular diamagnetic contribution of the equatorial region,
\begin{eqnarray}
   \frac{ \triangle E_{corr,1} }{\triangle N_{e,1}^{H = 0}} = \frac{1}{16\, r} \, \hbar \omega .
 \label{q18}
\end{eqnarray}

\begin{figure}
\resizebox{0.4\textwidth}{!} {
\includegraphics{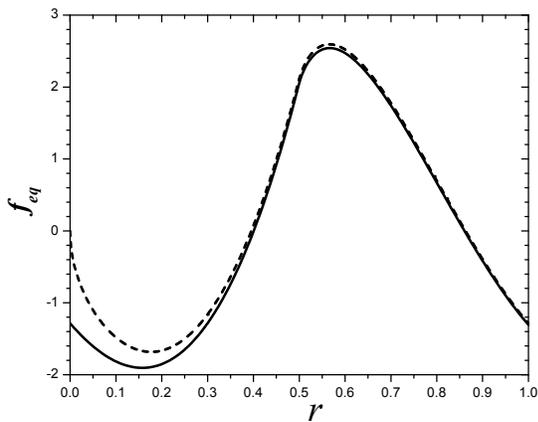}   }

\caption{
The oscillatory behavior of the irregular contribution to energy, $\triangle E_{eq} \sim f_{eq}(r)$,
from the equatorial region,
expressed in terms of the dimensionless parameter $r \sim 1/H$.
The dashed line stands for the contribution from the first equatorial tube, the solid line -- from the
first and second equatorial tubes.
} \label{fig0_7}
\end{figure}

Collecting all energy terms together,
we arrive at
\begin{eqnarray}
   \triangle E_{eq}= 2 \frac{m \pi v}{15 \sqrt{2}} \;
 \omega^2 \sqrt{\frac{m}{\hbar}\, \omega} \; f_{eq}(r) .
 \label{q21}
\end{eqnarray}
(The factor 2 stands for two equivalent contributions from the upper and lower Fermi semisphere.)
For the first equatorial tube we have $f_{eq}(r) = f_{eq,1}(r)$,
and the function $f_{eq,1}(r)$ has different dependences for cases $a$ and $b$, described earlier.
In case $a$ ($0 \le r < 1/2$)
$f_{eq,1}(r) = f_{eq,1}^a(r)$,
\begin{subequations}
\begin{eqnarray}
 f_{eq,1}^a(r) = \sqrt{r}\, (32 r^2 - 5) ,
 \label{q22a}
\end{eqnarray}
in case $b$ ($1/2 \le r < 1$) $f_{eq,1}(r) = f_{eq,1}^b(r)$,
\begin{eqnarray}
 f_{eq,1}^b(r) = \sqrt{r}\, (32 r^2 - 5 )- 80 \left(r-\frac{1}{2}\right)^{3/2} .
 \label{q22b}
\end{eqnarray}
\end{subequations}
The dependence of $\triangle E_{eq,1} \sim f_{eq,1}(r)$ from $r$ is shown in Fig.\ \ref{fig0_7}.
Note that $\triangle E_{eq,1}(r=0) \neq \triangle E_{eq,1}(r=1)$,
although $r=0$ and $r=1$ refer to the same physical situation.
Below we shall see that by including two equatorial tubes, the equality of the energy at $r=0$ and $r=1$
is restored.
%
%

\begin{figure}
\resizebox{0.4\textwidth}{!} {
\includegraphics{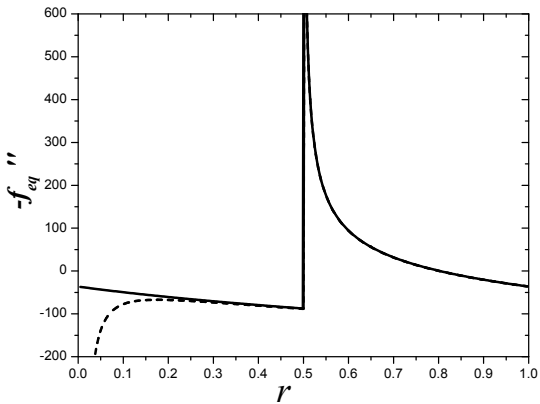}   }

\vspace{2mm}
\caption{
The oscillatory dependence of the magnetic susceptibility $\chi_{eq} \sim -f''_{eq}(r)$
from the equatorial region
expressed in terms of the dimensionless parameter $r \sim 1/H$ (see text).
The dashed line stands for the contribution from the first equatorial tube, the solid line -- from the
first and second equatorial tubes.
} \label{fig0_8}
\end{figure}

In calculating the magnetic susceptibility $\chi_{eq}$ one has to keep in mind that
$\triangle E_{eq}$ depends on $H$ through $\omega$ explicitly and on $r$ implicitly.
It can be shown that the contribution from the derivative of $r(H)$ with respect to the magnetic field $H$ is dominant.
Finally, we obtain
\begin{eqnarray}
  \chi_{eq} = -\frac{\sqrt{2} m \pi v}{15} \;
 \omega^2 \sqrt{\frac{m}{\hbar}\, \omega} \; \frac{\partial^2 f_{eq}(r) }{\partial r^2 } \left( \frac{\partial r}{\partial H} \right)^2 .
 \label{q24}
\end{eqnarray}
The plot of $\chi_{eq}(r)$ is reproduced in Fig.\ \ref{fig0_8}.
It is worth noting that $\chi_{eq,1}$ diverges at $r \rightarrow 0^+$
(the divergence disappears when the second equatorial tube is accounted for)
and at $r \rightarrow (1/2)^+$.
The latter persists in a more refined calculation with two or more equatorial tubes, because
it is connected with the onset of the occupation of a new Landau level in the $(k_x,k_y)$ equatorial plane.
%
%

Notice that if we limit ourselves to the case of only the first equatorial tube, then in correspondence with Eqs.\ (\ref{q22a}) and (\ref{q22b}),
the energy values at $r=0$ and $r=1$ are different, namely $\triangle E_{eq,1}(0) = 0$, $\triangle E_{eq,1}(1) \neq 0$,
Fig.\ \ref{fig0_7}.
In reality the physical situation is the same, the condition $r=0$ simply implies that the first equatorial tube is absent,
while the second equatorial tube plays the role of the first.
The inconsistence exists for the other quantities, for example, for the magnetic susceptibility, Fig.\ \ref{fig0_8}.
Therefore, to make the values at $r=0$ and $r=1$ consistent, we have to take into account the irregular term
from the second equatorial tube.
Then the contribution from the equatorial region $\triangle E_{eq}$, described by (\ref{q21}), changes,
\begin{eqnarray}
  \triangle E_{eq}=\triangle E_{eq,1}+\triangle E_{eq,2} .
 \label{ne0}
\end{eqnarray}
and the function $f_{eq}(r)$ in (\ref{q21}) becomes
\begin{eqnarray}
  f_{eq}(r)=f_{eq,1}(r)+f_{eq,2}(r) .
 \label{ne0a}
\end{eqnarray}
Numerical results for two equatorial tubes are shown
by solid lines in Figs.\ \ref{fig0_6}, \ref{fig0_7} and \ref{fig0_8}.
It is worth noting that except for the range around $r=0$ and $r=1$, the inclusion of the second equatorial tube
plays only a minor role.

\section{Monovalent metal}
\label{sec:realmet}

\subsection{Magnetic tubes and their properties}
\label{sec:tube}

In an externally applied magnetic field $\vec{H}$ pointing along the $z$-axis,
the energy spectrum of the occupied electron states of real metals is completely changed.
Here and below we follow the semiclassical representation of electron orbits in
the real and momentum ($\hbar \vec{k}$) space as briefly described in Sec.\ \ref{ssec:gast}.
It is worth mentioning that the semiclassical picture for itinerant electrons has been supported by the
equation of motion method of Ref.\ \cite{Rot2}, provided that in the quantization condition, Eq.\ (\ref{l1}),
$\gamma = 1/2 + h \Gamma_1 + h^2 \Gamma_3 \approx 1/2$
(here $h=eH/c$). Notice however, that in the limit of small magnetic field $h \rightarrow 0$
and $\gamma \rightarrow 1/2$. Therefore, in the following we will still use $\gamma = 1/2$, although this is not
critical for the method, see details in Appendix~\ref{subsec:gam}.

In the momentum space the $n$-th Landau orbit in the $(k_x, k_y)$ plane is defined by the area $A_n$ Eq.\ (\ref{l1b}),
and the orbital motion is associated with the energy $E_{\perp,n}$, Eq.\ (\ref{gg3}).
The real space orbit defined by Eq.\ (\ref{l1}) is no longer a circle, but a closed curve corresponding to the $\vec{k}-$orbit in the $(k_x, k_y)$ plane rotated
through $\pi/2$ about the field direction and scaled by $\hbar c/e H$.
In real metals, in contrast to the free electron gas, Eq.\ (\ref{g4}),
the cyclotron frequency $\omega$ is given by
\begin{eqnarray}
     \omega = \frac{eH}{c\, m^*} ,
\label{l4}
\end{eqnarray}
where the effective cyclotron mass $m^*$ is defined as
\begin{eqnarray}
     m^* = \frac{\hbar^2}{2 \pi} \frac{\partial A}{\partial E} .
\label{l4b}
\end{eqnarray}
Here $A$ is the area of the electron orbit in the $(k_x,k_y)$-plane of the $\vec{k}$-space.
In practice, $\partial A/\partial E$ is found from the contour integral
\begin{eqnarray}
      \frac{\partial A}{\partial E} = \oint \frac{dk}{\left| \left( \frac{\partial E}{\partial \vec{k}} \right)_{\perp}  \right|} .
\label{l4c}
\end{eqnarray}
From Eq.\ (\ref{l4c}) we conclude that in general for different Landau levels $n$ and $m$ with energies $E_n \neq E_m$
we have $(\partial A/\partial E)_n \neq (\partial A/\partial E)_m$ and consequently $m^*_n \neq m^*_m$, $\omega_n \neq \omega_m$.
For real metals for neighboring Landau levels like $n$ and $n' = n \pm 1$ since $\hbar \omega \ll E_F$ we still have
$m^*_n \approx m^*_m$, $\omega_n \approx \omega_m$.

So far, it has been implied that $k_z = 0$.
Since $k_z$ is a constant of the electron motion \cite{AM},
we can easily extent our consideration to the general case with $k_z \neq 0$ as has been done in Sec.\ \ref{ssec:gast}
for the free electron gas.
The most important difference with the Fermi gas is that the quantities $A_n$, $\partial A/\partial E$ and hence
$m^*$ and $\omega$ depend on $k_z$.
Thus, in general the total electron energy is given by
\begin{eqnarray}
     E = E_{\perp,n} + E_{z}(k_z) ,
\label{l0}
\end{eqnarray}
where $E_{z}(k_z)$ is the electron energy associated with the one dimensional (1D) electron band describing
the electron motion in the $z-$direction.
Notice that at $k_z = 0$ we have $E_z = 0$. Therefore, in general
$E_{z}(k_z) = E(\vec{k}) - E_{\perp,n}(\vec{k})$, provided that $\vec{k}$ lies on the Landau level $n$.
In the free electron case $E_{z}(k_z)=\hbar^2 k_z^2/2 m$.

As in the case of the Fermi gas,
we select in the $\vec{k}$-space a tube, whose main property will be that its number of electron states and energy of all states
do not change in the presence of the external magnetic field.
At each component $k_z$ we consider auxiliary quantized electron orbits $O_n$ with areas
\begin{eqnarray}
     A_n^{aux} = \frac{2\pi eH}{c \hbar} \, (n + \delta)
\label{l5b}
\end{eqnarray}
in the $(k_x, k_y)$ plane, where $\delta \sim H$ is a small parameter ($\delta \ll 1$) which is discussed in detail in Appendix~\ref{subsec:gam}.
Its choice is determined by imposing the equality of energy (see Eq.\ (\ref{l11b}) below) to a very high accuracy.
At fixed value of $H$, $A_n^{aux}$ is a constant area throughout $k_z$ although the shape of the orbit can vary with $k_z$.
We associate with $A_n^{aux}$ a certain energy of the electron movement in the $(k_x, k_y)$ plane, which we denote as $E_n^{aux}$.
At $k_z = 0$ this energy coincides with the orbital band energy, i.e. $E_n^{aux} = E(\vec{k})$ where
$\vec{k}$ lies on the $O_n$ orbit.
For $k_z \neq 0$ using the proximity to $E_{\perp,n}$ we have $E_{n}^{aux} \approx \hbar \omega (n + \delta)$, Appendix~\ref{subsec:gam}.

%
%
\begin{figure}
\resizebox{0.45\textwidth}{!} {
\includegraphics{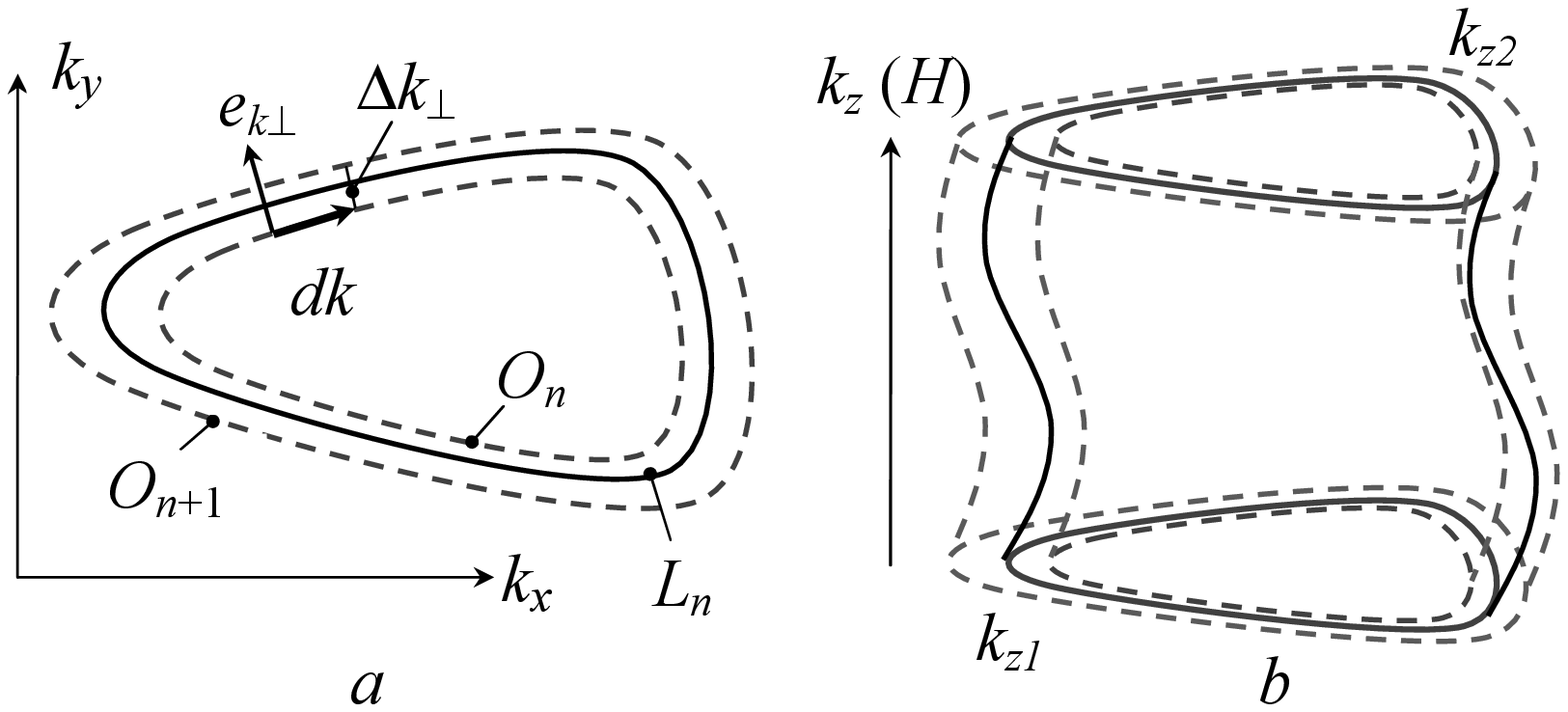}  }
%
\caption{Magnetic tube and the corresponding Landau level.
On the left ($a$): the $(k_x,k_y)$ tube cross-section and the $n$-th Landau level (the solid line $L_n$).
The dashed circles correspond to two auxiliary quantized orbits: $O_n$ and $O_{n+1}$, see text for details.
On the right ($b$): the tube in the $\vec{k}$-space, $k_{z1} \le k_z \le k_{z2}$.
Without magnetic field electron states are distributed throughout the whole tube, in the presence they are
condense on the Landau level in the middle.
}
 \label{fig1}
\end{figure}
Note that at any fixed $k_z$ the $n$-th Landau level $L_n$ whose area is given by Eq.\ (\ref{l1b}),
is sandwiched between auxiliary orbits $O_n$ and $O_{n+1}$, Eq.\ (\ref{l5b}), as shown in Fig.\ \ref{fig1}.
Moreover, for a band with dispersion the Landau energy $E_{\perp,n}$, Eqs.\ (\ref{gg3}), lies between $E_n^{aux}$ and $E_{n+1}^{aux}$.
Since $E_{n+1}^{aux} - E_{n}^{aux} \approx \hbar \omega \ll E_F$, within the interval $(A_n^{aux},\, A_{n+1}^{aux})$
between the auxiliary orbits $n$ and $n+1$ we deal with an infinitesimal situation.
Below we show that in the $(k_x, k_y)-$plane the number of electron states with energies $E_n^{aux} \leq E \leq E_{n+1}^{aux}$
without magnetic field
equals the number of electron states condensing on the $n$-th Landau level in the presence of the field.
The same holds for their total energy.
We start by noting that the density of electron states ${\cal N}_{\perp}(E)$ in the $(k_x,k_y)$-plane is
\begin{eqnarray}
     {\cal N}_{\perp}(E_{\perp}) = \frac{d N_{\perp} }{d E_{\perp} } = 2 \frac{L_x L_y}{(2\pi)^2} \frac{\partial A}{\partial E}
     = \frac{2 m^*}{\hbar^2} \frac{L_x L_y}{2\pi} .
\label{l9}
\end{eqnarray}
(Here the double spin degeneracy is taken into account.)
Using (\ref{l9}) and (\ref{l4b}), we find that the number of electron states in the $n$-th tube without field, i.e.
in the energy range $E_n^{aux} \leq E \leq E_{n+1}^{aux}$, is
\begin{eqnarray}
     \triangle N_n(H=0) = \int_{E_n^{aux}}^{E_{n+1}^{aux}} {\cal N}_{\perp}(E_{\perp}) dE_{\perp}    \nonumber \\
     = \frac{2 L_x L_y}{(2 \pi)^2} (A_{n+1}^{aux} - A_n^{aux}) = 2 N_p ,
\label{l10a}
\end{eqnarray}
where $N_p$ is the spatial degeneracy of the Landau levels, Eq.\ (\ref{l8}).
Calculating the energy of these states without field to the accuracy of $(\hbar \omega / E_F)^3 \triangle N_n$, Appendix~\ref{subsec:gam}, we obtain
\begin{eqnarray}
   {\cal E}_n(H=0) = \int_{E_n^{aux}}^{E_{n+1}^{aux}} {\cal N}_{\perp}(E_{\perp}) E_{\perp} dE_{\perp}
   = E_{\perp,n} \triangle N_n . \nonumber \\
\label{l10b}
\end{eqnarray}
That is, ${\cal E}_n(H=0)$ coincides with the energy of these states on the $n$-th Landau level in the presence of the field.
Thus, we have proven that
\begin{subequations}
\begin{eqnarray}
  & &   \triangle N_n(H=0) = \triangle N_n( H \neq 0) = 2 N_p , \label{l11a} \\
  & &   {\cal E}_n(H=0) = {\cal E}_n(H \neq 0) = \hbar \omega \left( n+\frac{1}{2} \right) 2 N_p . \quad \quad \label{l11b}
\end{eqnarray}
\end{subequations}

Eqs.\ (\ref{l11a}) and (\ref{l11b}) are valid for any $k_z-$component, and therefore for the whole tube $n$, Fig.\ \ref{fig1},
defined by its upper ($k_z^{(2)}$) and lower ($k_z^{(1)}$) boundaries, $k_z^{(1)} \leq k_z \leq k_z^{(2)}$.
As for the free electron gas, $k_z^{(2)}$ and $k_z^{(1)}$ are conveniently defined by the intersection of a tube with the Fermi surface, Fig.\ \ref{fig2}.
As a result, we have two $n$-th tubes: the first tube (denoted by $1$ in the inset of Fig.\ \ref{fig2}) lies entirely inside the Fermi surface and
does not exhibit diamagnetism, while
the second tube (denoted by $2$ in the inset of Fig.\ \ref{fig2}) is only partially filled and
results in a diamagnetic response.
We consider this effect in the following sections.

\subsection{Diamagnetically active electron states}
\label{sec:pert}

\begin{figure}
\resizebox{0.4\textwidth}{!} {
\includegraphics{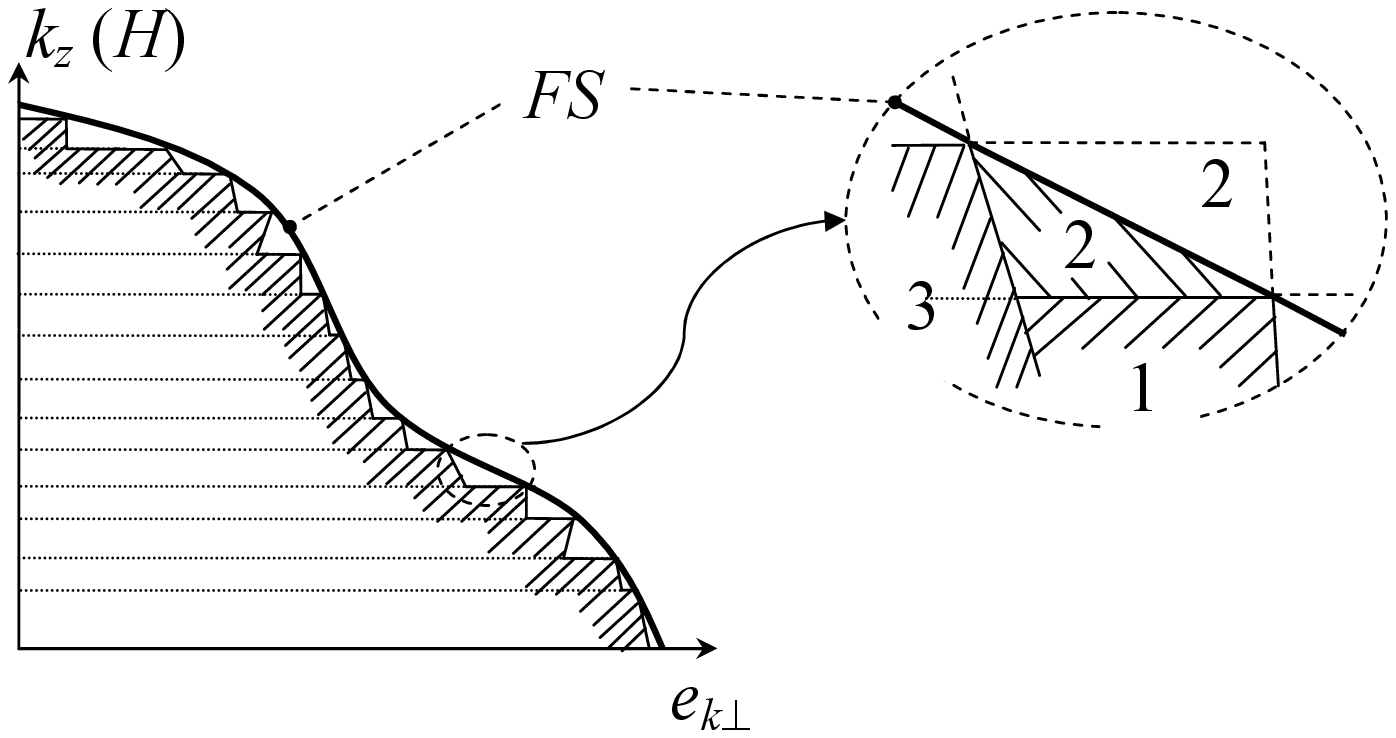}   }

\caption{
Magnetic tubes and the Fermi surface ($FS$) in a magnetic field in the $(k_z, e_{k_{\perp}})$-cross-section in the $\vec{k}$-space.
All electron states within the tubes inside the Fermi surface shown as the dashed area on the left, are completely filled,
and their full energy does not change in the magnetic field.
The diamagnetic effect is due to the electron states of partially occupied tube states (like the tube 2 in the inset),
making a step-like structure near the Fermi surface. Inset: 1 -- Completely occupied tube $n$,
2 -- partially occupied tube $n$, 3 -- completely occupied tube $n-1$.
} \label{fig2}
\end{figure}

Consider the Fermi surface and define magnetic tubes in an externally applied magnetic field $H$
pointing in the positive $z-$direction, Fig.\ \ref{fig2}.
As has been discussed in Sec.\ \ref{sec:tube}, each magnetic tube $n$
at $k_z$ is sandwiched between two boundary electron orbits $O_n$ and $O_{n+1}$ in the $(k_x, k_y)-$plane
whose areas $A_n^{aux}$ and $A_{n+1}^{aux}$ are defined by Eq.\ (\ref{l5b}).
As a result we obtain a set of boundary surfaces,
which intersect the Fermi surface in orbits perpendicular to the $z$-axis.
The intersection orbits unlike simple circles for the Fermi gas, Sec.\ \ref{sec:feg}, are rather complicated, and below we describe them in detail.
These orbits lie in $(k_x, k_y)-$planes and we can use them to define two $k_z-$boundary conditions, $k_z^{(n,1)}$ and $k_z^{(n,2)}$,
in such a way that the magnetic tube $n$ lies entirely inside the Fermi surface.
The fully occupied tubes are schematically shown in Fig.\ \ref{fig2}.
In Sec.\ \ref{sec:tube} we have seen that the electron states of the completely filled tubes do not change their energy in magnetic field.
Therefore, the whole effect is due to the states lying in the partially occupied tubes passing through the Fermi surface
(like tube 2 in the inset of Fig.\ \ref{fig2}).
%
%

\begin{figure}
\resizebox{0.4\textwidth}{!} {
\includegraphics{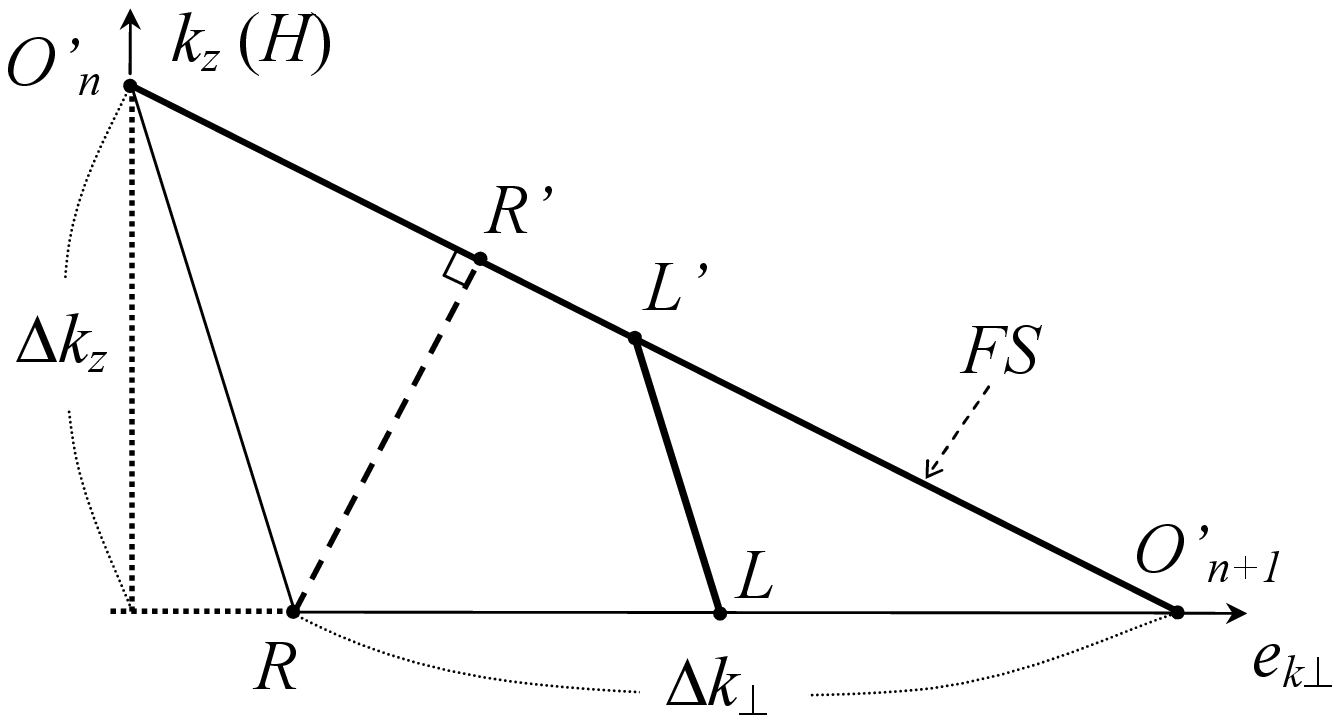}   }

\caption{
The $(k_z, e_{k_{\perp}})$-cross-section of a partially occupied tube $n$ near the Fermi surface.
$FS$ stands for the Fermi surface, $L L'$ for Landau orbits $L_n$, $O'_n$ and $O'_{n+1}$ for auxiliary
quantized orbits. $R$ is the energy reference point, $E_{ref}=E(R)=E_F - \hbar \omega$,
$E(L)=E_{ref} + \hbar \omega/2$, $\triangle l_F = |O'_n\, O'_{n+1}|$, $\triangle k_F = |R R'|$.
} \label{fig3}
\end{figure}

Consider a boundary orbit $O_{n,F}$ (defined by $A_n^{aux} = const$, Eq.\ (\ref{l5b})) in the $(k_x, k_y)-$plane lying on the Fermi surface.
It defines a certain value of $k_z$ which we denote by $k_{z,n}$. The electron energy band gradient, $\partial E(\vec{k}) / \partial \vec{k}$,
on $O_{n,F}$
is perpendicular to the Fermi surface and hence to the orbit. In addition, we will use the projection of the gradient in
the plane of the orbit, $(\partial E / \partial \vec{k})_{\perp}$, which is also perpendicular to $O_{n,F}$, Figs.\ \ref{fig1}, \ref{fig3}.
The normalized vector in the direction of $(\partial E / \partial \vec{k})_{\perp}$ will be denoted by $e_{k_{\perp}}$.
Notice that plane $(k_z, e_{k_{\perp}})$ is normal to $O_{n,F}$. Cross-sections of magnetic tubes passing through the Fermi surface
are schematically shown by horizontal dotted lines in Fig.\ \ref{fig2}.
(In general, Fig.\ \ref{fig2} should be understood as composed of many different panels -- at most one panel for each tube cross-section,
because the energy gradient $\partial E / \partial \vec{k}$ and hence $(k_z, e_{k_{\perp}})$-planes still can have different orientations.)

Consider a typical partially occupied tube $n$, whose triangle cross-section in the $(k_z, e_{k_{\perp}})$-plane is shown in Fig.\ \ref{fig3}.
Points $O'_n$ and $O'_{n+1}$ are defined by the intersection of the $(k_z, e_{k_{\perp}})$-plane with the orbits $O_{n,F}$ and $O_{n+1,F}$, respectively.
The side $O'_n,O'_{n+1}$ is the intersection with the Fermi surface, the side $O'_n, R$ is the intersection
with the $A_n^{aux}$ boundary surface, and the side $O'_{n+1}, R$ is the intersection with the $(k_x,k_y)$-plane of the orbit $O_{n+1,F}$.
It is convenient to count energy from $E_{ref} = E(R) = E_F - \hbar \omega$. The vector $\vec{R R'}$ being perpendicular to $O'_n,O'_{n+1}$
is collinear with the energy gradient $\partial E / \partial \vec{k}$. Finally, the line $L,L'$ is given by the intersection
of Landau orbits $L_n$ at various $k_z$, Eq.\ (\ref{l1b}), with the $(k_z, e_{k_{\perp}})$-plane.
Since we consider the limit of small magnetic fields, i.e. $\hbar \omega / E_F \rightarrow 0$,
the triangle is assumed infinitesimal allowing for linear dependencies of relevant quantities.
In particular, the line $L,L'$ is parallel to $R, O'_n$, etc., and we can take $\delta = 0$ in Eq.\ (\ref{l5b}), $\gamma = 1/2$ in Eq.\ (\ref{l1}),
Appendix~\ref{subsec:gam}.
%
%

In the following we will need $\triangle k_{\perp}$, $\triangle k_{z}$, $\triangle k_F$ expressed in terms of
$(\partial E / \partial \vec{k})_F$, $(\partial E / \partial \vec{k}_{\perp})_F$, $(\partial A/\partial k_z)_F$
calculated at the Fermi surface. Details of these calculations are given in Appendix~\ref{subsec:st}.

Now in accordance with Eq.\ (\ref{g11}) we can find the number of the active electron states in all partially occupied tubes.
We recall that in Eq.\ (\ref{g11}) $\triangle N_n \equiv \triangle N(k_z,\; k_z + \triangle k_z)$ is the number of the electron states
in the $n$-th partially filled tube, whose $k_z-$component lies between $k_{z,n} \equiv k_z$ and $k_{z,n+1} \equiv k_z + \triangle k_z$.
The calculation of $\triangle N$ carried out in Appendix \ref{subsec:st}, results in Eq.\ (\ref{g10a}), that is
\begin{eqnarray}
  \frac{ \triangle N}{\triangle k_z } = 2\pi v \frac{e H}{\hbar c} = const . \label{f10a}
\end{eqnarray}
Notice that the right hand side of Eq.\ (\ref{f10a}) is independent of the effective mass $m^*$
and $\partial A / \partial E$.
It should be emphasized that the polar and equatorial regions of the Fermi surface
are special cases, which are discussed in Sec.~\ref{realdHvA}.
It turns out that the polar regions fully comply with the general expression (\ref{f10a}), while
the equatorial regions (where $\partial A(k_z) / \partial k_z = 0$) give rise to the well known
de Haas-van Alphen oscillations of the magnetic susceptibility,
which is not the subject of the present study.

For small magnetic fields when $\triangle k_z \ll k_F$,
in Eq.\ (\ref{g11}) we can substitute the summation with the integration, Eq.\ (\ref{g12a}).
In Eq.\ (\ref{g12a}) $k_z^{min}$ ($k_z^{max}$) is the minimal (maximal) $k_z-$component of the vectors $\vec{k}$ lying on the Fermi surface.
Using (\ref{f10a}) [or Eq.\ (\ref{g10a})] we arrive at Eq.\ (\ref{f15}),
where $\ell_z \equiv k_z^{max} - k_z^{min}$ is defined exclusively by the projection of the Fermi surface along the direction of the applied magnetic field.
Notice however, that unlike for free electron gas, now in general  $k_z^{min} \neq -k_F$, $k_z^{max} \neq k_F$.
Furthermore, as discussed later in Sec.\ \ref{sec:real} in real metals $\ell_z$ depends on the orientation of the applied magnetic field
with respect to the Fermi surface.

Thus, the number of active electron states $N \sim H$, their total energy change in the magnetic field $\sim H^2$,
which gives a constant diamagnetic susceptibility $\chi$.

\subsection{Landau diamagnetic susceptibility}
\label{sec:Lan}

%
\begin{figure}
\resizebox{0.45\textwidth}{!} {
\includegraphics{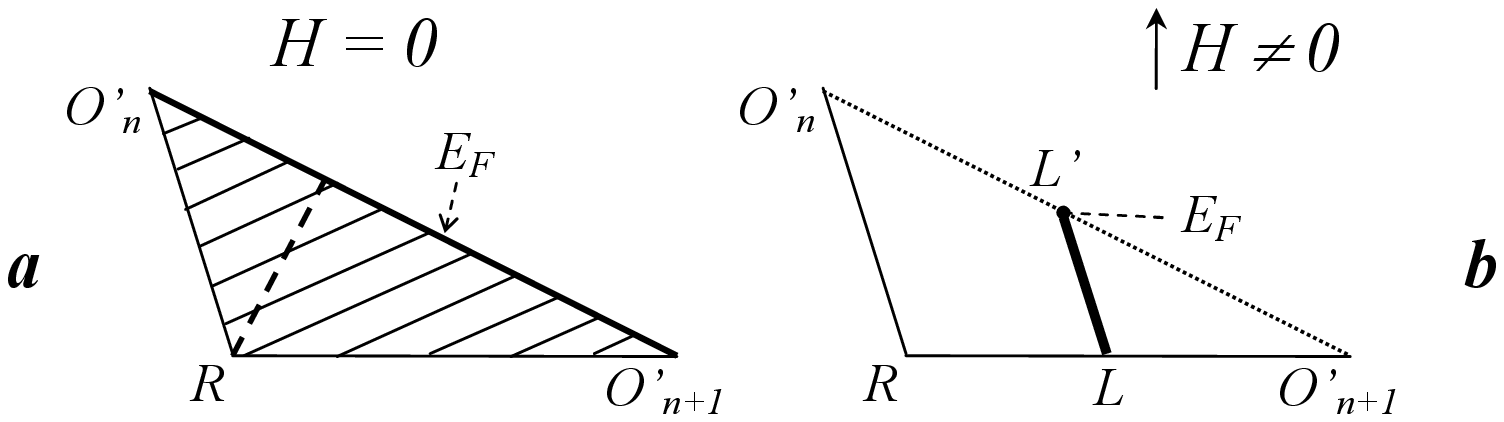}   }

\vspace{2mm}
\caption{
The $(k_z, e_{k_{\perp}})$-cross-section of a partially occupied tube $n$ near the Fermi level.
On the left ($a$): occupied electron states without magnetic field, $\vec{H}=0$.
On the right ($b$): the occupied electron states on Landau levels (the $LL'$ bold line) in the magnetic field $\vec{H} \neq 0$.
$E_F$ is the Fermi energy of the one-dimensional electron states, whose transverse energy $E_n$
is determined by the $n$-th Landau energy, $E_n = E_{n,\perp}(L)$.
} \label{fig4}
\end{figure}
%
The transformation of the electron states of a partially occupied tube is schematically shown in Fig.\ \ref{fig4}.
In the applied magnetic field $H \neq 0$ all electrons of the partially occupied tube are on the $n$-th Landau orbits with the transverse energy $E_n$,
and occupy the lowest in energy $k_z$-states, as shown in Fig.\ \ref{fig4}.

As we have seen in Sec.\ \ref{ssec:gasd} for the free electron gas the highest occupied energy in the presence of
the field coincides with the Fermi energy $E_F$ obtained in the absence of the field.
For the real metal the $(k_z, e_{k_{\perp}})$-cross-section of a partially occupied tube $n$, Figs.\ \ref{fig3} and \ref{fig4},
in general is very different from the $(k_z,k_x)$ cross-section of the Fermi gas shown in Figs.\ \ref{fig0_3} and \ref{fig0_4}.
However, the highest occupied energy is still $E_F$. This follows from Eq.\ (\ref{e12}), written for the
partially occupied tube $n$ for real metal.
In terms of the $z-$projection $\delta k_z$ of the wave vector $\vec{k}_{L'}$ shown in Fig.\ \ref{fig3},
it implies $\delta k_z = \triangle k_z / 2$ just as for the Fermi gas, Eq.\ (\ref{eg12a}).
Therefore, there are no electron transitions between different tubes and
we can calculate the energy change for each tube separately and then sum all contribution for
the final result.

Keeping in mind that in the magnetic field there are two energy contributions:
$E_{\perp}$ in the $(k_x,k_y)$ plane and $E_z$ at $k_z$, Eq.\ (\ref{l0}),
we now calculate the energy change for the tube $n$ in the applied magnetic field, Eq.\ (\ref{eg2n}).

The calculation of all components is performed in Appendix~\ref{subsec:en} (with respect to the energy $E_{ref}$, Fig.\ \ref{fig4}),
and results in Eqs.\ (\ref{eg10}), (\ref{eg11a}) and (\ref{eg11b}). For the total energy change
we get $\triangle E = \hbar \omega\, \triangle N / 12$.
This is the same expression as for the free electron gas, Eq.\ (\ref{eg18}), but now the cyclotron frequency $\omega$
is different for different tubes, Eq.\ (\ref{l4}).
Therefore, by means of Eqs.\ (\ref{l4}) and (\ref{f10a}) we rewrite the expression for $\triangle E$ in terms of $\triangle k_z$:
\begin{eqnarray}
  \triangle E = \frac{1}{6} \pi v \left( \frac{eH}{c} \right)^2 \frac{\triangle k_z}{m^*} .
 \label{ee18b}
\end{eqnarray}

Note that Eq.\ (\ref{ee18b}) refers to any partially filled tube.
Therefore, making the summation over all tubes
we find
\begin{eqnarray}
  E = \frac{1}{48} \frac{V}{\pi^2} \left( \frac{eH}{c} \right)^2 \int_{k_z^{min}}^{k_z^{max}} \frac{d k_z}{m^*(k_z)} .
 \label{e19}
\end{eqnarray}
Here, as before $k_z^{min}$ ($k_z^{max}$) is the minimal (maximal) $k_z-$component of the Fermi surface.
Notice, that $m^*$ found from Eqs.\ (\ref{l4b}) and (\ref{l4c}) for orbits in the $(k_x,k_y)$ plane lying on the Fermi surface,
 is a function of $k_z$.
Finally, for the magnetic susceptibility $\chi$ we have:
\begin{eqnarray}
  \chi = -\frac{d^2E(H)}{dH^2} = - V \frac{1}{24} \frac{e^2}{\pi^2 c^2} \int_{k_z^{min}}^{k_z^{max}} \frac{d k_z}{m^*(k_z)} . \nonumber \\
 \label{e20}
\end{eqnarray}
In the case of the free electron gas the integral on the right hand side of (\ref{e20}) is reduced to $\ell_z/m = 2 k_F/m$ and we arrive at
Eq.\ (\ref{g20}) for the diamagnetic susceptibility of the Fermi gas.
In general, Eq.\ (\ref{e20}) differs from the pioneer expression of Peierls \cite{Pei}, Eq.\ (\ref{i2}), although for the case of parabolic bands
in terms of $k_x$ and $k_y$, Eq.\ (\ref{e2}), they give the identical results, Appendix~\ref{subsec:Pei}.

As a final and important remark we note that the density of electron states, $g(E_F)$, remains unchanged on applying the magnetic field,
although energy gradients at the Fermi surface change their directions and values.
We prove this remarkable effect in Appendix~\ref{subsec:dos}.

\subsection{Special cases and extremal cross-sections of the Fermi surface}
\label{realdHvA}

General expressions for $\triangle k_{\perp}$ and $\triangle k_z$, Eqs.\ (\ref{a1}) and (\ref{a2}), become indefinite
when $(\partial E / \partial \vec{k} )_{\perp} = 0$ or $\partial A / \partial k_z = 0$.
The condition $(\partial E / \partial \vec{k} )_{\perp} = 0$ occurs at polar points of the Fermi surface and the full account of this situation
is given in Appendix \ref{subsec:polar}. In that case there are no deviations from the general final equations (\ref{f10a}) and (\ref{ee18b}).

If $\partial A(k_z) / \partial k_z = 0$ then
from Eq.\ (\ref{a2}) it follows that $\triangle k_z \rightarrow \infty$, and such an extremal cross-section
corresponds to a maximum or a minimum of $A(k_z)$ as a function of $k_z$.
It is well known that extremal orbits and extremal regions give rise to the oscillatory
behavior for the diamagnetic response (de Haas -- van Alphen effect), Ref.\ \cite{Sho}.
In our approach for the Fermi gas described in detail in Sec.\ \ref{sec:dHvA},
it has also been demonstrated that the extremal region consisting of a few tubes in the equatorial region
of the Fermi surface, results in the oscillatory behaviour of the diamagnetic susceptibility.
This should occur in real metals as well, but
in this paper we consider only the regular contribution to energy ($\triangle E_{eq}^L$) leaving the irregular part ($\triangle E_{eq}^{irr}$)
for future consideration.

\section{Application to alkali metals}
\label{sec:real}

We have applied the method to calculations of the diamagnetic response of alkali metals: Li, Na, K, Rb and Cs,
which are crystallized in the body centered cubic (bcc) lattice and have only one active electron band.
As follows from Sec.\ \ref{sec:realmet} for diamagnetic susceptibility one needs two very important characteristics:
the Fermi surface and the energy gradient $\nabla_{\vec{k}}E$ on it.
Therefore, we first have performed {\it ab initio} density functional calculations (DFT) of the alkali metals
using Moscow-FLAPW code \cite{lapw}.
The Perdew-Burke-Ernzerhof (PBE) \cite {PBE} variant of the generalized gradient approximation has been employed,
with the number of $k$-points 2470.
For the Fermi surface and gradient calculations we have used the tetrahedron method \cite{tet}.
In this method the irreducible part of the Brillouin zone is divided into a set of tetrahedra and
within each tetrahedron the linear dependence of the band energy $E(\vec{k})$ is assumed.
The cross-sections of the constant energy $E=E_F$ in the $\vec{k}$-space from all relevant tetrahedra give the pieces forming the Fermi surface.
Also, in each tetrahedron crossing the Fermi surface we find the value of the energy gradient, $\nabla_{\vec{k}}E$,
which is unchanged within the tetrahedron.

\begin{figure}[!]
\resizebox{0.45\textwidth}{!} {
\includegraphics{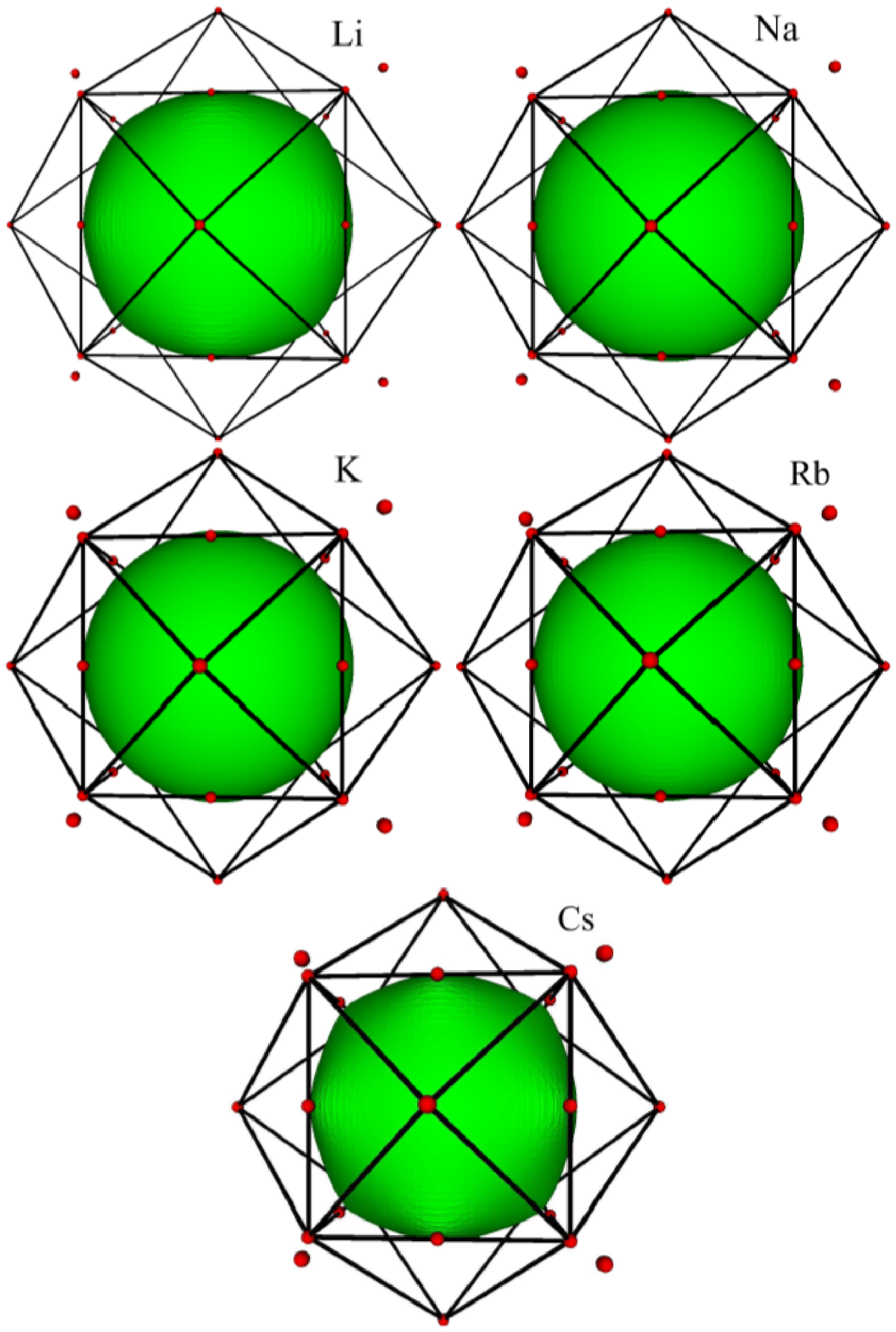}   }


\caption{Calculated Fermi surfaces of alkali metals.} \label{fig5}
\end{figure}

The obtained Fermi surfaces (FS) are shown in Fig.\ \ref{fig5}.
The important property of the calculated Fermi surfaces is that the topological space inside each of them is simply (or path) connected.
In other words, the occupied electron states form a single piece Fermi surface.
(The application of the presented method to more complicated Fermi surfaces and to completely filled electron bands
requires additional considerations.)
%

%
\begin{figure}[!]
\resizebox{0.45\textwidth}{!} {
\includegraphics{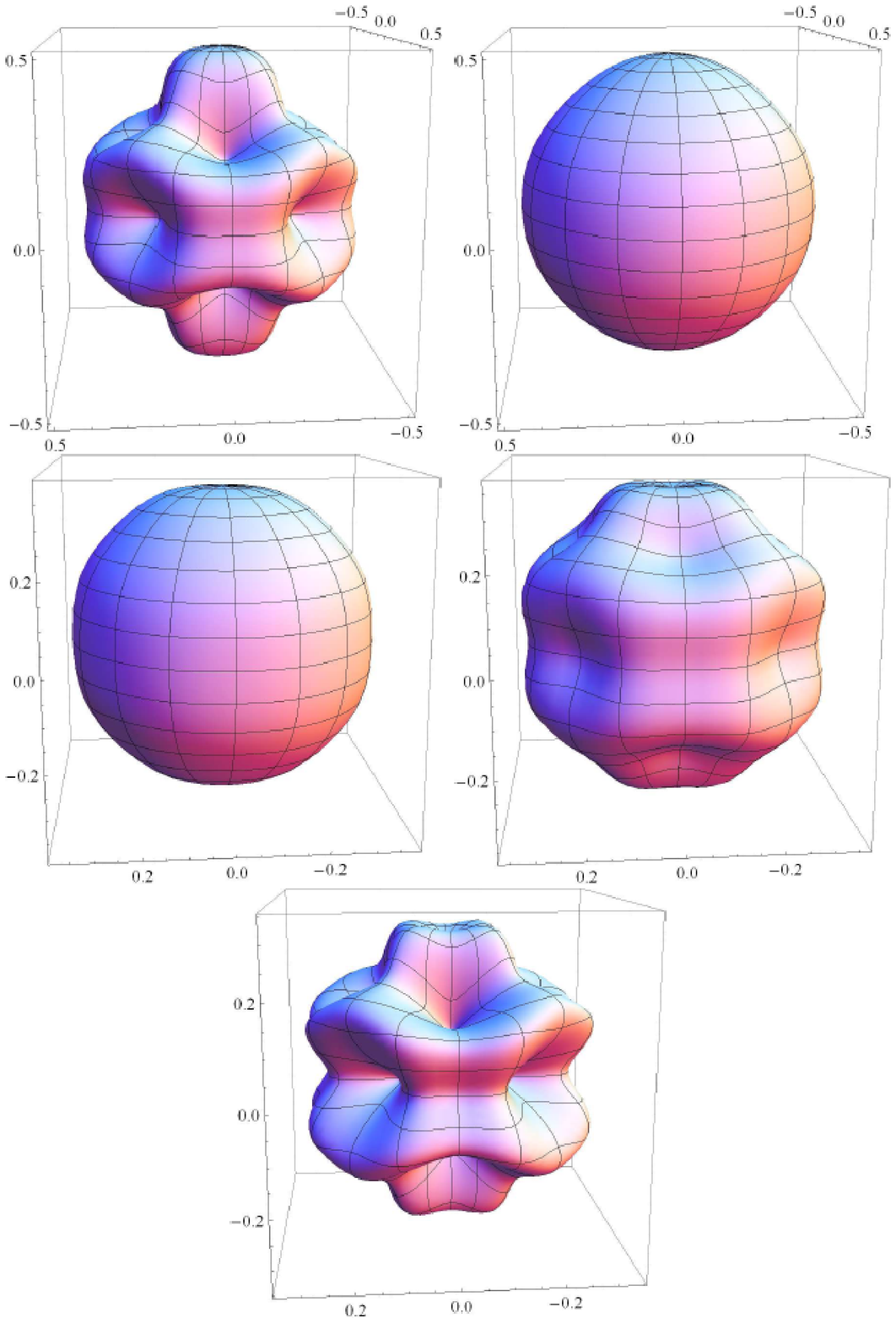}   }


\caption{Calculated energy gradient modula of alkali metals. (The order of the elements as in Fig.\ \ref{fig5}.)} \label{fig6}
\end{figure}

Since the Fermi surface is invariant under all symmetry operations of the crystal (in our case we deal with the $O_h$ cubic symmetry),
it can be expanded in terms of symmetry adapted functions (SAFs) belonging to the fully symmetric ($A_{1g}$) representation of $O_h$ \cite{BC}.
Such SAFs known as cubic harmonics \cite{BC}, are linear combinations of real spherical harmonics $Y_l^{m,\tau}$
[where $\tau=c$, $s$ stands for the azimuthal dependence of the cosine ($c$) or sine ($s$) type] with $l = 4$, 6, 8,... :
\begin{eqnarray}
  & &K_4(\theta,\phi) = \sqrt{\frac{7}{12}} Y_4^0 + \sqrt{\frac{5}{12}} Y_4^{4,c}, \nonumber \\
  & &K_6(\theta,\phi) = \sqrt{\frac{1}{8}} Y_6^0 - \sqrt{\frac{7}{8}} Y_6^{4,c}, \nonumber \\
  & &K_8(\theta,\phi) = \sqrt{\frac{33}{64}} Y_8^0 + \sqrt{\frac{7}{48}} Y_8^{4,c} + \sqrt{\frac{65}{192}} Y_8^{8,c} . \nonumber
 \label{r1}
\end{eqnarray}
Here $(\theta,\phi) = \hat{k}$ are polar angles in $k-$space and for spherical harmonics we use the phase convention of Bradley and Cracknell \cite{BC}.
(For compactness here and below we omit $(\theta,\phi)$ in $Y_l^{m,\tau}$.)
Therefore, the Fermi surface can be very accurately represented by the expansion
\begin{eqnarray}
  k_F(\hat{k}) &=& C_0^{FS} Y_0^0 + C_4^{FS} K_4(\hat{k}) \nonumber \\
  & &+ C_6^{FS} K_6(\hat{k}) + C_8^{FS} K_8(\hat{k}) ,
 \label{r2a}
\end{eqnarray}
where $Y_0^0 = 1/\sqrt{4\pi}$ is the zeroth spherical harmonic ($l=m=0$).
In the expansion (\ref{r2a}) the Fermi surface is completely defined by the coefficients $C_0^{FS}$, $C_4^{FS}$, $C_6^{FS}$ and $C_8^{FS}$.
These coefficients can be obtained numerically from the calculated Fermi surfaces, Fig.\ \ref{fig5}.
For all alkalis they are given in Table \ref{tab1}.
%
\begin{table}
\caption{Calculated characteristics of alkali metals: $\chi_L$ is the steady diamagnetic susceptibility
($av$ stands for the averaged value),
$\chi_P$ is the Pauli paramagnetic susceptibility, $m^*$ is the effective cyclotron mass, $k_H^{max}$ is
the maximal projection of the Fermi surface on the direction of the applied magnetic field $\vec{H}$,
$\gamma_{el}$ is the linear electron specific heat coefficient (in mJ/mole K$^2$),
$N(E_F)$ is the electron density of states (in 1/eV). $C_l^{FS}$ ($l=0$, 4, 6, 8)
are the coefficients describing the shape of the Fermi surface and $C_l^{g}$ are the coefficients of the expansion of
the modulus of the energy gradient ($\vec{\nabla}_k E_F$). $a$ is the bcc lattice constant (in {\AA}),
and indices $[001]$, $[101]$ or $[111]$ define
the direction of $\vec{H}$, at which the corresponding quantity is calculated, see text for details.
\label{tab1} }

\begin{ruledtabular}
\begin{tabular}{c | c  c  c  c  c }

metal                     & Li       & Na     &  K     &   Rb     &  Cs     \\
\tableline
   $a$, {\AA}             & 3.50     & 4.29   &  5.35  &  5.585 &  6.14   \\
$\chi_L^{av}$, $10^{-7}$  & -1.753   & -2.086 & -1.624 & -1.515 & -1.179 \\
 $\chi_L[001]$, $10^{-7}$ & -1.670   & -2.081 & -1.631 & -1.508 & -1.142 \\
 $\chi_L[101]$, $10^{-7}$ & -1.871   & -2.090 & -1.631 & -1.560 & -1.308 \\
 $\chi_L[111]$, $10^{-7}$ & -1.714   & -2.084 & -1.618 & -1.485 & -1.114 \\

   $m^*_{av}$             &  1.517   & 1.036  & 1.067  & 1.097  & 1.300 \\

    $m^*_{av}[001]$       &  1.552   & 1.038  & 1.065  & 1.099  & 1.310 \\
    $m^*_{av}[101]$       &  1.462   & 1.035  & 1.065  & 1.080  & 1.214 \\
    $m^*_{av}[111]$       &  1.527   & 1.037  & 1.070  & 1.107  & 1.317 \\

    $k_H^{max}[001]$      & 0.576    & 0.480  & 0.386  & 0.369  & 0.333 \\
    $k_H^{max}[101]$      & 0.609    & 0.481  & 0.386  & 0.375  & 0.353 \\
    $k_H^{max}[111]$      & 0.582    & 0.481  & 0.385  & 0.366  & 0.326 \\
    $N(E_F)$              & 0.486    & 0.494  & 0.792  & 0.888  & 1.302 \\
 $\gamma_{el}$            & 1.146 & 1.166 & 1.867  & 2.094  & 3.068 \\
    $\chi_P$, $10^{-7}$   & 12.207   & 6.728  & 5.565  & 5.487  & 6.051 \\

    $C_0^{FS}$          & 2.08862  &  1.70433 &  1.36665 &  1.30909 &  1.19032 \\
    $C_4^{FS}$          & -0.01529 & -0.00039 &  0.00092 & -0.00048 & -0.00474 \\
    $C_6^{FS}$          & -0.02358 & -0.00065 & -0.00101 & -0.00694 & -0.02149 \\
    $C_8^{FS}$          & 0.00660  & -0.00001 &  0.00074 &  0.00263 &  0.00914 \\

    $C_0^{g}$           & 1.40005  &  1.64429 &  1.28035 &  1.19546 &  0.94458 \\
    $C_4^{g}$           & 0.13752  &  0.01058 & -0.00908 &  0.00294 &  0.04106 \\
    $C_6^{g}$           & 0.28278  &  0.01325 &  0.00441 &  0.08695 &  0.24620 \\
    $C_8^{g}$           & -0.10748 & -0.00267 & -0.00733 & -0.05090 & -0.12383 \\

\end{tabular}
\end{ruledtabular}
\end{table}

The other quantity -- energy gradient at the Fermi surface $\vec{\nabla}_k E_F(\hat{k})$ -- also can be expanded in terms of the cubic harmonics, Eq.\ (\ref{r1}).
Since the gradient is always normal to the Fermi surface and therefore its direction can be found at any point of the Fermi surface
we need only its modulus expansion,
\begin{eqnarray}
  \left| \vec{\nabla}_k E_F(\hat{k}) \right| = C_0^{g} Y_0^0 + C_4^{g} K_4(\hat{k}) \nonumber \\
  + C_6^{g} K_6(\hat{k}) + C_8^{g} K_8(\hat{k}) .
 \label{r2b}
\end{eqnarray}
The coefficients $C_0^{g}$, $C_4^{g}$, $C_6^{g}$ and $C_8^{g}$ obtained from the {\it ab initio} calculations are also quoted in Table \ref{tab1}.
The three dimensional picture of $\left| \vec{\nabla}_k E_F(\hat{k}) \right|$ as a function of
the direction $\hat{k}$ in $k-$space is shown in Fig.\ \ref{fig6}.

From Fig.\ \ref{fig5} and the values of the coefficients $C_l^{FS}$ we conclude that the Fermi surfaces of lithium and cesium demonstrate largest deformations from the spherical shape.
Sodium and potassium, on the other hand,
have FS shapes which are very close to spherical, while deviations for rubidium lie between these two groups.
This finding is even more clearly illustrated by the shape of energy gradients, Fig.\ \ref{fig6}, which are more sensitive to deviations from the spherical form.

\begin{figure}[!]
\resizebox{0.4\textwidth}{!}
{\includegraphics{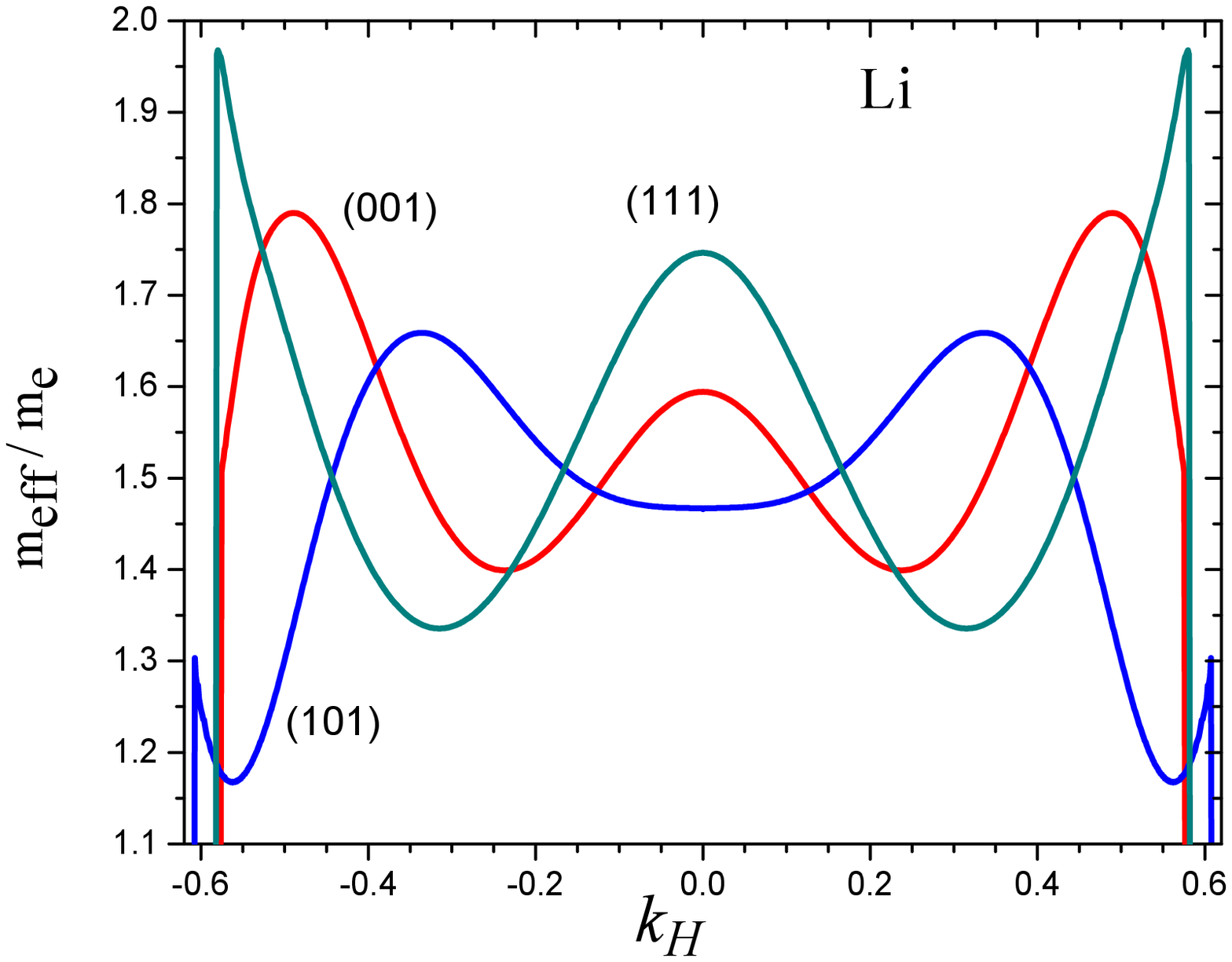}}
\resizebox{0.4\textwidth}{!}
{\includegraphics{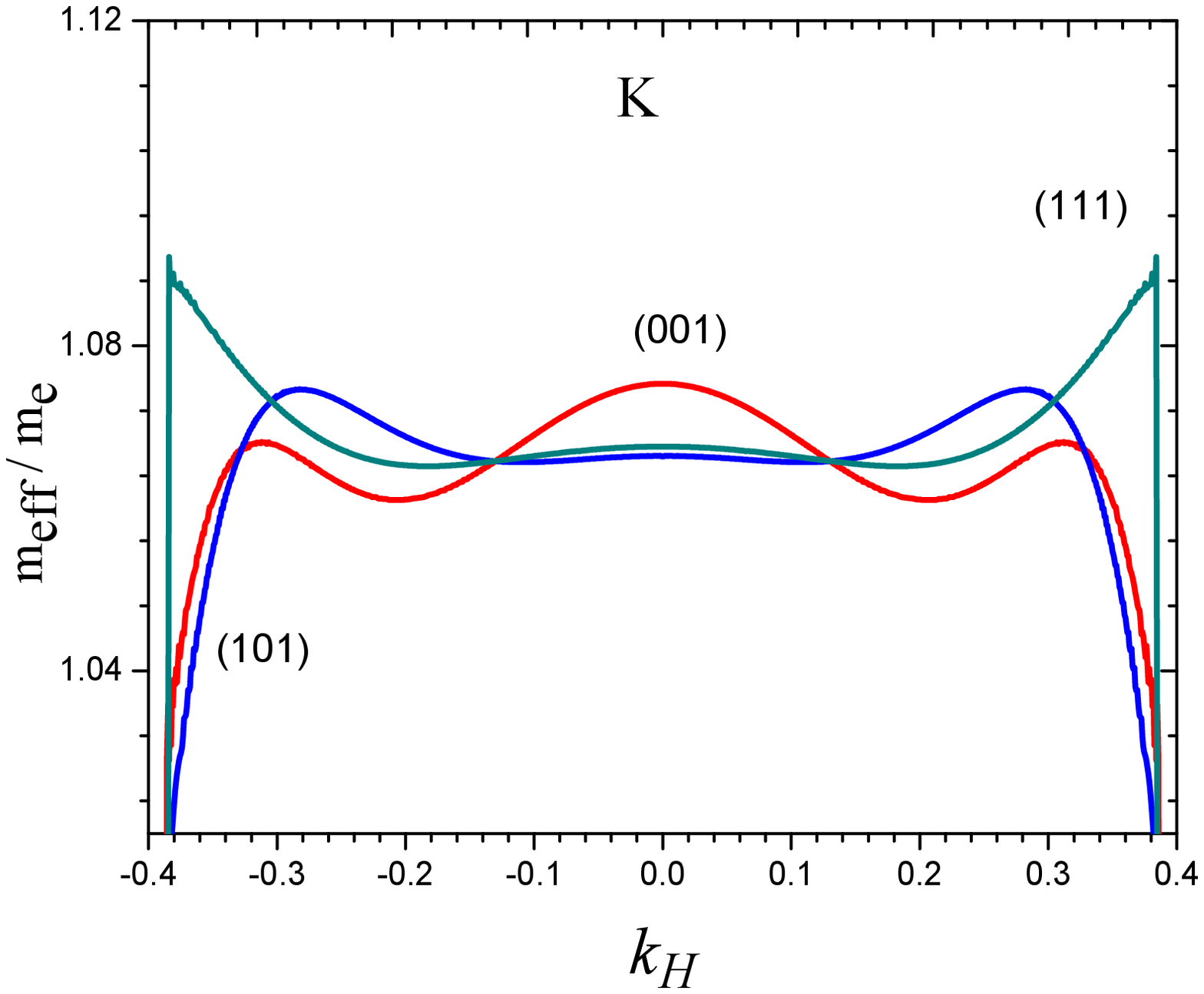}}

\caption{Calculated cyclotron mass $m^*$ for Li and K
for directions $[001]$, $[101]$ and $[111]$ of the magnetic field $H$.} \label{fig7}
\end{figure}

We now study the cyclotron mass $m^*$, which depends on the closed orbit on the Fermi surface in $k$-space,  Eqs.\ (\ref{l4b}), (\ref{l4c}).
The orbit is given by the cross section of the Fermi surface with a plane lying perpendicular to the applied magnetic field $\vec{H}$.
Such a plane is uniquely defined by its projection $k_H \equiv k_{z'}$ on the $z'$-axis parallel to $\vec{H}$.  Therefore,
the cyclotron mass in the applied magnetic field is a function of $k_H$ and $\hat{H}$ pointing in the direction of $\vec{H}$,
that is $m^*= m^*(\hat{H},\, k_H)$.
In Fig.\ \ref{fig7} we plot $m^*$ as a function of $k_H$ for three directions of magnetic field for
potassium with nearly spherical Fermi surface and lithium with a deformed Fermi surface.
%

Notice that for deformed Fermi surfaces (lithium and cesium) $k_H^{max}$ and $k_H^{min} = -k_H^{max}$
are noticeably different for various directions of $H$, Table \ref{tab1}.
In Table \ref{tab1} we also quote average cyclotron masses ($m^*_{av}$) for magnetic field directions $[001]$, $[101]$ and $[111]$,
defined according to the following relation,
\begin{eqnarray}
  \frac{1}{m^*_{av}} = \frac{1}{|k_H^{max} - k_H^{min}|} \int_{k_H^{min}}^{k_H^{max}} \frac{d k_H}{m^*(k_H)} .
 \label{r3}
\end{eqnarray}

Since in general $m^*_{av}$ depends on $\hat{H}$, the Landau diamagnetic susceptibility $\chi_L$ defined by Eq.\ (\ref{e20}), also shows this dependence.
Therefore, to obtain the averaged value $\chi_L^{av}$ for each element we perform an effective integration of $\chi_L(\hat{H})$
over $\hat{H}$, using a special set of 170 points \cite{Leb}. (The set gives the correct coefficients of expansion up to
spherical harmonics with $l = 10$.)
The calculated averaged values of $\chi_L^{av}$ and $m^*_{av}$ are given in Table \ref{tab1}.
They are in fair correspondence with estimations of the Landau contributions obtained with other approaches \cite{Oli}.

\begin{figure}[!]
\resizebox{0.45\textwidth}{!}
{\includegraphics{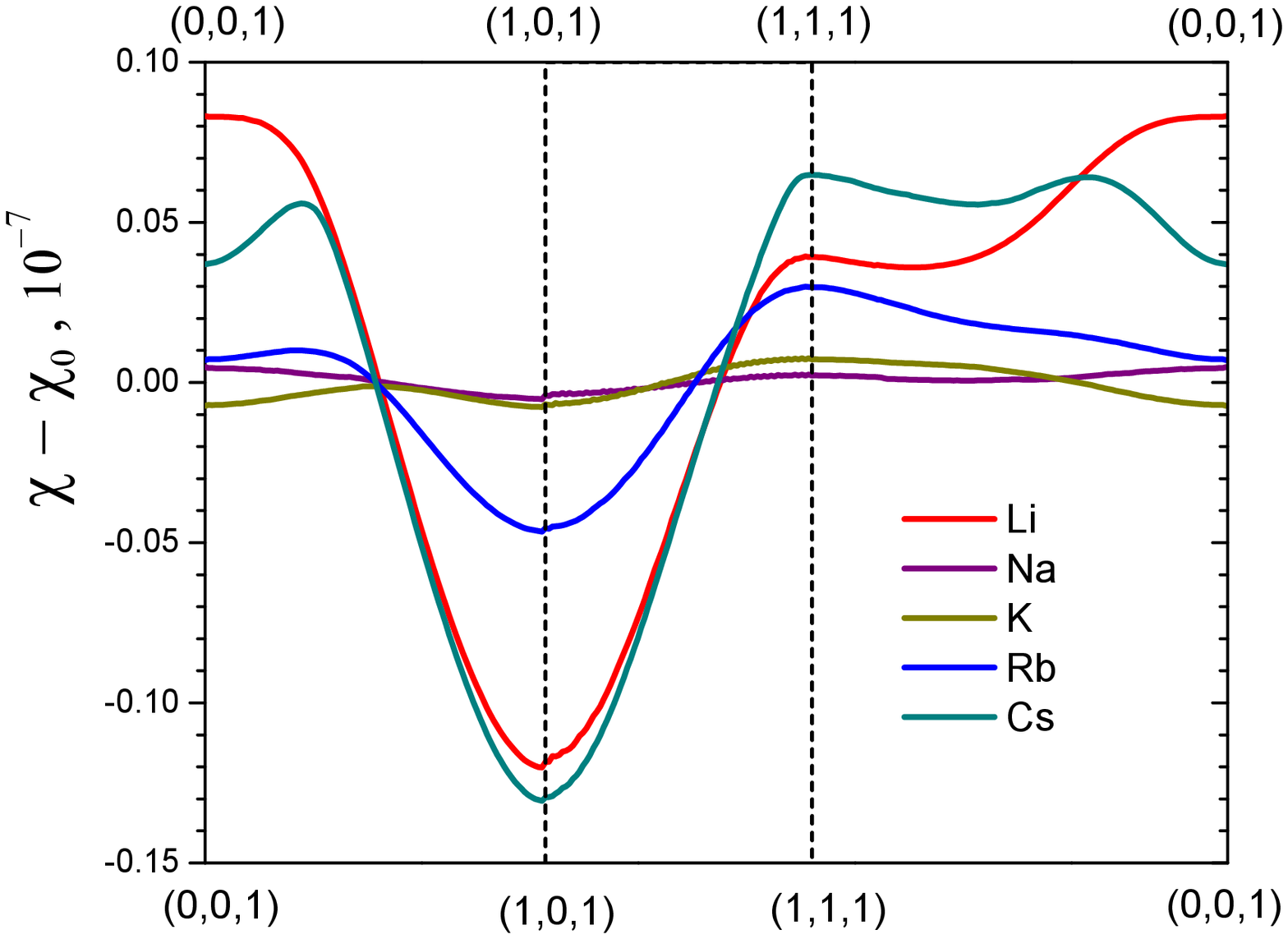}}

\caption{Calculated deviations of the Landau diamagnetic susceptibility $\chi_L$
from its average value $\chi_L^{av}$ for various directions
of the applied magnetic field $H$ for alkali metals.} \label{fig8}
\end{figure}

For a chosen direction of magnetic field $\hat{H}$, $\chi_L$ in general differs from the average value $\chi_L^{av}$, and we can study
$\chi_L - \chi_L^{av}$ as a function of $\hat{H}$. These dependencies
for all alkali metals are presented in Fig.\ \ref{fig8}, where in addition to the selected directions ($[001]$, $[101]$ and $[111]$)
we give data for intermediate angles (i.e. lying on the circumferences fusing $[001]$ and $[101]$, $[101]$ and $[111]$, $[111]$ and $[001]$).
%
Notice that the deviations from $\chi_L^{av}$ are larger for lithium and cesium and smaller for sodium and potassium.
The anisotropy of $\chi_L$ with respect to the direction of the applied magnetic field is a remarkable property of
the Landau diamagnetism, which singles it out from the Pauli paramagnetism and the Langevin diamagnetism of closed electron shells.

\section{Conclusions}
\label{sec:con}

The steady diamagnetic response for the Fermi gas and a real metal with a general simply connected Fermi surface is obtained
analytically at zero temperature ($T = 0$) within the widely accepted semiclassical approach \cite{Ons,Lif,Sho,semi,Rot2}.
The diamagnetic effect is caused by electron states in a very narrow region of the Fermi surface.
The consideration is based on a structure in $\vec{k}$-space called a magnetic tube,
which sandwiches the Landau level inside it.
The completely occupied tubes are diamagnetically inert. Only partially occupied tubes located at the Fermi surface
are responsible for the diamagnetic effect.
Although energy gradients at the Fermi surface change in the magnetic field, the total density of electron states at the Fermi energy
remains constant, Appendix~\ref{subsec:dos}.

We have applied the method for calculations of the Landau diamagnetic susceptibility $\chi_L$ of alkali metals (Li, Na, K, Rb, Cs).
The crucial quantities for that are the Fermi surface $k_F(\hat{k})$ and energy gradients $\vec{\nabla}_k E_F(\hat{k})$ on it,
which are obtained from {\it ab initio} band structure calculations.
For accurate calculations of $\chi_L$, we expand both $k_F(\hat{k})$ and $|\vec{\nabla}_k E_F(\hat{k})|$
in terms of symmetry adapted functions, Eqs.\ (\ref{r2a}) and (\ref{r2b}).
We have demonstrated that the diamagnetic effect depends on the direction $\hat{H}$ of the applied magnetic field $\vec{H}$.
The anisotropy of $\chi_L$ is larger for lithium and cesium and smaller for sodium and potassium.
It is worth noting that the Langevin diamagnetism of atomic closed core shells and the Pauli paramagnetism
are isotropic with respect to $\hat{H}$.

The present approach can be applied to other metals or intermetallic compounds.
The method can also be extended to the case of non-zero temperatures, where the thermal excitations
of one dimensional electron gas consisting of electron states on the Landau levels should be taken into account.

While the steady response is due to the region just below the Fermi surface, the oscillatory behavior of
energy and magnetic susceptibility arises from a few partially occupied tubes located at its extremal cross-sections.
A small oscillatory change of the Fermi energy of the free electron gas in the applied magnetic field
is caused by a transfer (inflow or outflow) of electrons from this equatorial region
of the Fermi surface, Sec.\ \ref{sec:dHvA}.


\appendix

\section{Calculation of $\triangle k_{\perp}$, $\triangle k_z$, $\triangle N$}
\label{subsec:st}

Here we quote some useful relations in a partially occupied tube shown in Fig.\ \ref{fig3},
taking into account the infinitesimal properties of its triangular $(k_z, e_{k_{\perp}})$ cross-section.
In particular, we have
\begin{eqnarray}
  \triangle k_{\perp} = \frac{\hbar \omega}{\left| ( \partial E / \partial \vec{k} )_{\perp} \right|} > 0.
 \label{a1}
\end{eqnarray}
To calculate $\triangle k_z$ we first consider the area $A$ of the $(k_x, k_y)$-cross-section of the Fermi surface
as a function of $k_z$, that is, $A(k_z)$.
Then $\triangle A = A_{n+1}^{aux} - A_{n}^{aux} = (\partial A(k_z)/ \partial k_z) \triangle k_z$, and we arrive at
\begin{eqnarray}
  \triangle k_z =  \frac{2 \pi e H}{c \hbar} \frac{1}{|\partial A / \partial k_z |} > 0 .
 \label{a2}
\end{eqnarray}

For the number of the electron states in the partially occupied tube $n$, Fig.\ \ref{fig3}, we find
\begin{eqnarray}
  \triangle N =  2 v \oint_{O_{n+1}} \triangle S dk = v \triangle k_z \oint_{O_{n+1}} \triangle k_{\perp} dk  \nonumber \\
  = v\, \triangle A \, \triangle k_z,
 \label{a3}
\end{eqnarray}
where $v = V /(2\pi)^3$ and $\triangle S$ is the area of the $(k_z, e_{k_{\perp}})$-cross-section:
$\triangle S = \triangle k_{\perp} \triangle k_z / 2$.
This gives Eqs.\ (\ref{g10a}) and (\ref{f10a}).

\section{Calculation of $\triangle E^{H = 0}$, $\triangle E_{\perp}^{H \neq 0}$, $\triangle E_z^{H \neq 0}$}

\label{subsec:en}

We first consider the energy of a partially occupied tube $n$ without magnetic field ($H = 0$), left panel of Fig.\ \ref{fig4}.
The reference energy at $\vec{k}(R)$ is $E_{ref} = E(\vec{k}(R)) = E_F - \hbar \omega$, Fig.\ \ref{fig3}.
We will calculate the energy by integrating over $x$ from $R$ to $R'$ along the energy gradient direction $\partial E / \partial \vec{k}|_F$ , Fig.\ \ref{fig3}.
Notice that $O'_n O'_{n+1}$ is perpendicular to $R R'$, the energy at $x$ is $(x / \triangle k_F) \hbar \omega$ and the infinitesimal volume element
is $dk\, dx (x / \triangle k_F) |O'_n, O'_{n+1}|$.
Denoting $\triangle l_F = |O'_n, O'_{n+1}|$, we get
\begin{eqnarray}
 \triangle E^{H = 0} = 2 v \oint_{O_{n+1}} \epsilon\, dk ,
 \label{c1}
\end{eqnarray}
where
\begin{eqnarray}
 \epsilon = \int_0^{\triangle k_F} \left( \frac{x}{\triangle k_F} \hbar \omega \right) \left( \frac{x}{\triangle k_F} \triangle l_F \right) dx .
 \label{c2}
\end{eqnarray}
Performing integration in (\ref{c2}) and
taking into account that the area of the cross-section shown in Fig.\ \ref{fig3} is $\triangle S = \triangle k_F \triangle l_F /2$
whereas $2 v \oint \triangle S dk = \triangle N$, we arrive at Eq.\ (\ref{eg10}).

We now consider the case $H \neq 0$, the right panel of Fig.\ \ref{fig4}.
The transverse component of an electron state condensed on the Landau level is $E(L) = E_{ref} + \hbar \omega /2$.
Thus, for each electron state we have $\triangle E_{\perp}^{H \neq 0} = \hbar \omega /2$, and for all electron states we obtain Eq.\ (\ref{eg11a}).

Populating the Landau levels along the $L, L'$ line shown in Fig.\ \ref{fig3}, the one dimensional electron energy
changes from $E_{ref}$ to $E_{ref} + \hbar \omega /2$. Since in the infinitesimal cross-section approximation of Fig.\ \ref{fig3}
the energy increases linearly with $k_z$ and the density of electron states along the $L, L'$ line is constant,
for the energy of all these occupied states we obtain $\triangle E_z^{H \neq 0} = (\hbar \omega /4) \, \triangle N$, Eq.\ (\ref{eg11b}).

\section{Polar region}

\label{subsec:polar}

If the energy gradient at a certain point $P$ of the Fermi surface is parallel to the $z-$axis (or to the magnetic field $\vec{H}$),
that is $\partial E(P)/ \partial \vec{k}\; ||\, \vec{H}$,
then $(\partial E(P)/ \partial \vec{k})_{\perp} = 0$, but on the other hand, the contour path for $E=E(P)$ in Eq.\ (\ref{l4c}) reduces to the point $P$.
This leads to the ambiguity for $\partial A/\partial E$ in Eq.\ (\ref{l4c}) and consequently for $m^*$ and $\omega$ in Eqs.\ (\ref{l4b}) and (\ref{l4}).
This situation corresponds to a polar region around the point $P$ and requires a special treatment.

For convenience we put the origin of the coordinate system at the point $P$ and assume that the center of curvature is located below.
The Fermi surface in the vicinity of $P$ is given by the function $k_z^{FS}(k_x,k_y) = f(k_x, k_y)$,
whose tensor of the second derivatives is diagonalized:
\begin{eqnarray}
 \left. \frac{\partial^2 k_z^{FS}}{\partial k_x ^2} \right|_P < 0 , \quad  \left. \frac{\partial^2 k_z^{FS}}{\partial k_y ^2} \right|_P < 0 ,
  \quad \left. \frac{\partial^2 k_z^{FS}}{\partial k_x \partial k_y} \right|_P = 0 . \quad
 \label{b1}
\end{eqnarray}

This polar region of the Fermi surface is described by the function
\begin{eqnarray}
  k_z^{FS}(k_x,k_y) = \frac{1}{2} \left. \frac{\partial^2 k_z^{FS}}{\partial k_x ^2} \right|_P k_x^2
  + \frac{1}{2} \left. \frac{\partial^2 k_z^{FS}}{\partial k_y ^2} \right|_P k_y^2  .
 \label{b2}
\end{eqnarray}
In the polar region we deal with the zeroth magnetic tube, i.e. $n = 0$, and with the zeroth Landau level.
Two tube's auxiliary boundary conditions are reduced to
\begin{subequations}
\begin{eqnarray}
   & & A_{n=0}^{aux} = 0 ,   \\
   & & A_{1}^{aux} = \frac{2\pi eH}{\hbar c} . \label{b3b}
 \label{b3}
\end{eqnarray}
\end{subequations}
(Notice that here $\delta = 0$.)
Thus, the first quantized auxiliary orbit $O_{n=1}$ at the Fermi surface is an ellipse at $\triangle k_z = k_{z,1} < 0$,
which can be found by equating its area to $A_{1}^{aux}$, Eq.\ (\ref{b3b}).
The result is
\begin{eqnarray}
  \triangle k_z = -\frac{e H}{\hbar c}
  \sqrt{ \left. \frac{\partial^2 k_z^{FS}}{\partial k_x ^2} \right|_P \left. \frac{\partial^2 k_z^{FS}}{\partial k_y ^2} \right|_P  } .
 \label{b4}
\end{eqnarray}
Therefore, the whole polar region with $0 \ge k_z \ge \triangle k_z$ represents the zeroth partially occupied tube.
It can be shown that the volume of the polar region is
\begin{eqnarray}
  V_P = \frac{1}{2} |\triangle k_z| A_{1}^{aux} .
 \label{b5}
\end{eqnarray}
The same expression holds for a small sphere cap in the case of the Fermi sphere of the free electron gas.
The number of active electron states ($\vec{H}=0$) is then
\begin{eqnarray}
  \triangle N_{n=0} = 2 v V_P = |\triangle k_z| \frac{2\pi eH}{\hbar c} v .
 \label{b6}
\end{eqnarray}
The latter equation demonstrates that $\triangle N_0 / |\triangle k_z|$ follows the general relation (\ref{f10a})
although the curvature of the Fermi surface polar region has been explicitly taken into account.
Analogously, one can show that the other quantities (energy etc.) also follow the general consideration.

\section{Tuning auxiliary boundary conditions}
\label{subsec:gam}

In this section we will give details of derivation of Eqs.\ (\ref{l10b}) and (\ref{l11b})
and define the small parameter $\delta \sim H$ in Eq.\ (\ref{l5b}).
The middle part of Eq.\ (\ref{l10b}) can be integrated by parts,
\begin{eqnarray}
  {\cal E}_n(H=0) = 2 \frac{L_x L_y}{(2 \pi)^2}
  \left[ E_{n+1}^{aux} A_{n+1}^{aux} - E_{n}^{aux} A_{n}^{aux} \right.    \nonumber \\
  - \left. \int_{E_{n}^{aux}}^{E_{n+1}^{aux}} A(E)\, dE \right] .
 \label{g1}
\end{eqnarray}
Introducing notations $\triangle E = E_{n+1}^{aux} - E_{n}^{aux}$,
$\triangle A = (2 \pi e/c \hbar) H$, Eq.\ (\ref{g1}) can be written as
\begin{eqnarray}
  {\cal E}_n =
  \left[ E_{n+1}^{aux} (1 - \triangle^* + \delta) + E_{n}^{aux} (\triangle^* - \delta) \right] \triangle N_n ,
 \label{g2}
\end{eqnarray}
where
\begin{eqnarray}
  \triangle^* = \frac{1}{\triangle A\, \triangle E} \int_{E_{n}^{aux}}^{E_{n+1}^{aux}} A(E)\, dE\; - \; n .
 \label{g3}
\end{eqnarray}
We further expand $A(E)$ up to the second derivative term,
\begin{eqnarray}
  A(E) = A(E_0) + A' (E - E_0) + \frac{1}{2} A'' (E - E_0)^2,
 \label{g4}
\end{eqnarray}
where $E_0 = (E_{n+1}^{aux} - E_{n}^{aux})/2$ and we use short notations
$A' = \partial A(E_0) / \partial E$, $A'' = \partial^2 A(E_0) / \partial E^2$.
With Eq.\ (\ref{g4}) we obtain
\begin{eqnarray}
  \triangle^* = \frac{1}{2} + \delta - \frac{A''}{12} \frac{(\triangle E)^2}{\triangle A} + O(H^3) ,
 \label{g5}
\end{eqnarray}
where $O(H^3) \sim H^3$ and Eq.\ (\ref{g2}) becomes
\begin{eqnarray}
  {\cal E}_n(H=0) = \left[ E_0 + \frac{A''}{12} \frac{(\triangle E)^2}{\triangle A} \right]\, \triangle N_n .
 \label{g6}
\end{eqnarray}
On the other hand, for the energy of the tube in the magnetic field with respect to $E_0$ one has
\begin{eqnarray}
  {\cal E}_n(H \neq 0) = \left[ E_0 - \frac{\triangle A}{A'} \delta + \frac{1}{8} \frac{A''}{A'} (\triangle E)^2 \right]\, 2N_p .
 \label{g7}
\end{eqnarray}
Equating Eqs.\ (\ref{g6}) and (\ref{g7}), we obtain
\begin{eqnarray}
  \delta = \frac{\pi e}{12 c \hbar} \frac{A''}{(A')^2}\, H .
 \label{g8}
\end{eqnarray}
If, as in Eq.\ (\ref{l1}) one uses $\gamma = 1/2 + \Gamma_1 h$ \cite{Rot2}, then in Eq.\ (\ref{g8})
$\delta$ should be replaced with $(\delta - \gamma) \sim H$.
From Eq.\ (\ref{g8}) it follows that $\delta$ is also a function of energy, $\delta = \delta(E)$.
This in turn leads to a more complicated final expression,
\begin{eqnarray}
  \delta(E) = \frac{\pi e}{12 c \hbar}\, \frac{f(E)}{(1 + E f'(E)/f(E))}\, H ,
 \label{g9}
\end{eqnarray}
which can be solved iteratively, starting with $f_0(E)=A''(E)/A'(E)^2$ as in Eq.\ (\ref{g8}).

Quantities $\gamma$ and $\delta$ should be taken into account for the definition of the boundaries of the
occupied magnetic tubes, because they concern a large number of electron states inside the Fermi surface.
For partially occupied tubes at the Fermi surface they give only small
corrections of the order of $\hbar \omega/E_F$ to main results, which can be omitted.

\section{Connection with Peierls' expression}
\label{subsec:Pei}

In his early pioneer work \cite{Pei}, Peierls obtained Eq.\ (\ref{i2}) for the steady diamagnetic susceptibility.
Below we will show that the magnetic susceptibilities given by Eqs.\ (\ref{i2}) and (\ref{e20}) coincide for the
case of the parabolic energy band dependence in terms of $k_x$ and $k_y$,
\begin{eqnarray}
 E(\vec{k}) = a(k_z) k_x^2 + b(k_z) k_y^2 + c(k_z) k_x k_y + d(k_z) .
 \label{e2}
\end{eqnarray}
Notice that $a(k_z)$, $b(k_z)$, $c(k_z)$, $d(k_z)$ here are arbitrary smooth functions of $k_z$ and
the $z$-axis points in the direction of the applied magnetic field $H$.
The expression in the curly bracket of Eq.\ (\ref{i2}), being $\lambda = 4ab - c$, is an invariant of the quadratic form (\ref{e2}).
We then transform (\ref{e2}) to a new coordinate system $(k'_x,k'_y)$ where it takes the diagonal form
$E(\vec{k}) = a'(k_z) {k'_x}^2 + b'(k_z) {k'_y}^2 + d(k_z)$. Now $\lambda = 4 a' b'$ and in the following
will work in this representation.
For the integral in Eq.\ (\ref{i2}) we obtain
\begin{eqnarray}
 I = -4 \int_{FS} \frac{ a'(k_z) b'(k_z)\, dS(\vec{k})}{|\nabla_k E(\vec{k})|}    \nonumber  \\
 =  4 \pi \int_{k_z^{min}}^{k_z^{max}} \sqrt{a'(k_z) b'(k_z)}\, dk_z ,  \label{e3}
\end{eqnarray}
where the first integration is taken over the Fermi surface ($FS$) and
$k_z^{max}$ ($k_z^{min}$) is the maximal (minimal) projection of FS on the $z-$axis.
The final expression for Eq.\ (\ref{i2}) becomes
\begin{eqnarray}
  \chi = -\frac{e^2}{12 \pi^2 \hbar^2 c^2} \int_{k_z^{min}}^{k_z^{max}} \sqrt{a'(k_z) b'(k_z)}\, dk_z .
 \label{e4}
\end{eqnarray}

On the other hand, using Eq.\ (\ref{l4b}) we obtain
$m^*(k_z) = \hbar^2 /2 \sqrt{a'(k_z) b'(k_z)}$ .
Substitution of $m^*(k_z)$ in Eq.\ (\ref{e20}) gives the same result as before, Eq.\ (\ref{e4}).

In general however Eq.\ (\ref{i2}) differs from Eq.\ (\ref{e20}).
We ascribe it to a number of approximations used in the derivation of Eq.\ (\ref{i2}) \cite{Ada,KK,Bri}.
\\

\section{Density of electron states at $E_F$}
\label{subsec:dos}

Consider the partitioning of the Fermi surface region in partially occupied tubes.
The density of electron states (DOS) at the Fermi energy then is given by
\begin{eqnarray}
  g(E_F) \frac{1}{V} = \sum \triangle g_n(E_F) = \int_{k_{z_min}}^{k_{z,max}} dk_z \, \frac{\triangle g(E_F)}{\triangle k_z} , \nonumber \\
 \label{d1}
\end{eqnarray}
where
$\triangle g_n(E_F)$ is the DOS contribution from the part of the Fermi surface $S_n$ associated with a partially occupied tube $n$,
\begin{eqnarray}
  \triangle g_n(E_F) = \frac{1}{4 \pi^3} \oint_{S_n} \frac{dS}{|\vec{\nabla} E|} .
 \label{d2}
\end{eqnarray}
In the applied magnetic field $H \neq 0$, Fig.\ \ref{fig3}, electron states occupy the Landau level $n$ along the $L,L'$ line.
Since the Landau orbits are perpendicular to the $z-$axis, the energy gradient should be taken in the $z-$direction
with the surface element $dS$ in the $(k_x,k_y)-$plane.
From Fig.\ \ref{fig3} we obtain $\partial E / \partial k_z = \hbar \omega / \triangle k_z$, $dS_{\perp} = \triangle k_{\perp} dk $.
We thus arrive at
\begin{eqnarray}
  \triangle g^{H \neq 0}(E_F) = \frac{1}{4 \pi^3} \oint \frac{\triangle k_{\perp} \triangle k_z}{\hbar \omega} dk .
 \label{d3}
\end{eqnarray}
Notice that $\triangle S = \triangle k_{\perp} \triangle k_z /2$ is the area of the triangular ($O'_n, A, O'_{n+1}$) cross-section,
which can also be written as $\triangle S = \triangle k_{F} \triangle l_F /2$.
Since $|\partial E / \partial \vec{k}| = \hbar \omega / \triangle k_F$, we continue Eq.\ (\ref{d3}) as
\begin{eqnarray}
  \triangle g^{H \neq 0}(E_F) = \frac{1}{4 \pi^3} \oint \frac{\triangle l_F\, dk}{|\partial E / \partial \vec{k}|} = \triangle g^{H = 0}(E_F) . \nonumber \\
 \label{d4}
\end{eqnarray}
Here in establishing the last equality we have taken into account that $\triangle l_F\, dk = dS^{H=0}$ is a part of the Fermi surface perpendicular to
the energy gradient $\partial E / \partial \vec{k}$ in the absence of the magnetic field.
Eq.\ (\ref{d4}) leads to $g^{H \neq 0}(E_F) = g^{H=0}(E_F)$.


\end{document}